\documentclass[twocolumn,trackchanges]{aastex63}

\usepackage{natbib}
\usepackage{float}
\usepackage{amssymb}
\usepackage{amsmath}
\usepackage{verbatim}

\newcommand{\Rnum}[1]{\uppercase\expandafter{\romannumeral #1\relax}}

\newcommand{\NII}{[\mbox{N\,\textsc{ii}}]}

\newcommand{\SII}{[\mbox{S\,\textsc{ii}}]}

\newcommand{\Ha}{H$\alpha$}     
\newcommand{\ha}{H$\alpha$}   
\newcommand{\Hb}{H$\beta$}  
\newcommand{\hb}{H$\beta$}

\newcommand{\civ}{\mbox{C\,\textsc{iv}}}

\newcommand{\Mbh}{M\raisebox{-.3ex}{$\bullet$}}

\newcommand{\msun}{M$_{\odot}$}

\newcommand{\ergs}{erg s$^{-1}$}

\newcommand{\valerrud}[3]{${#1}^{+#2}_{-#3}$}
\newcommand{\verrud}[3]{{#1}^{+#2}_{-#3}}

\newcommand{\hr}{\mathrm{hr}}
\newcommand{\lumcont}{{\lambda L_\lambda \mathrm{(5100\,\AA)}}}
\newcommand{\lumct}[1]{{\lambda L_{\lambda;{#1}} \mathrm{(5100\,\AA)}}}
\newcommand{\jav}{\texttt{JAVELIN}}
\newcommand{\zdcf}{{$z$DCF}}
\newcommand{\swift}{\emph{Swift}}
\newcommand{\fvar}{{F_\mathrm{var}}}

\newcommand{\snu}{\affil{Department of Physics \& Astronomy, Seoul National University, Seoul 08826, Republic of Korea}}
\newcommand{\umich}{\affil{Department of Astronomy, University of Michigan, Ann Arbor, MI 48109, USA}}
\newcommand{\kasi}{\affil{Korea Astronomy and Space Science Institute, Daejeon 34055, Republic of Korea}}
\newcommand{\yonsei}{\affil{Department of Astronomy, Yonsei University, Seoul 03722, Republic of Korea}}
\newcommand{\nysc}{\affil{National Youth Space Center, Goheung 59567, Republic of Korea}}
\newcommand{\nasa}{\affil{NASA/GSFC, Code 662, Greenbelt, MD 20771, USA}}
\newcommand{\umd}{\affil{Department of Astronomy, University of Maryland, College Park, MD 20742, USA}}
\newcommand{\hefei}{\affil{Department of Astronomy, University of Science and Technology of China, Hefei 230026, China}}
\newcommand{\sandiego}{\affil{San Diego State University, San Diego, CA 92182, USA}}
\newcommand{\sai}{\affil{Sternberg Astronomical Institute, M.V. Lomonosov Moscow State University, Universitetski pr. 13, 119234 Moscow, Russia}}
\newcommand{\telaviv}{\affil{School of Physics and Astronomy and Wise Observatory, Tel-Aviv University, Tel-Aviv 6997801, Israel}}
\newcommand{\berkeley}{\affil{Department of Astronomy, University of California, Berkeley, CA 94720-3411, USA}}
\newcommand{\miller}{\affil{Miller Senior Fellow, Miller Institute for Basic Research in Science, University of California, Berkeley, CA  94720, USA}}

\newcommand{\byu}{\affil{Department of Physics and Astronomy, N283 ESC, Brigham Young University, Provo, UT 84602, USA}}
\newcommand{\belgrade}{\affil{Department of Astronomy, Faculty of Mathematics, University of Belgrade Studentski trg 16, Belgrade, 11000, Serbia}}
\newcommand{\fsu}{\affil{Department of Physics, Florida State University, Tallahassee, FL 32306, USA}}
\newcommand{\aob}{\affil{Astronomical Observatory Belgrade; Volgina 7; 11000 Belgrade, Serbia}}
\newcommand{\knu}{\affil{Major in Astronomy and Atmospheric Sciences, Kyungpook National University, Daegu 41566, Korea}}

\begin{document}

\title{Variability and the Size-Luminosity Relation of the Intermediate-Mass Active Galactic Nucleus in NGC 4395}

\author{Hojin Cho} \snu
\author{Jong-Hak Woo} \snu
\author{Edmund Hodges-Kluck} \umich \umd \nasa
\author{Donghoon Son} \snu
\author{Jaejin Shin} \snu \knu
\author{Elena Gallo} \umich
\author{Hyun-Jin Bae} \snu \yonsei
\author{Thomas G. Brink} \berkeley
\author{Wanjin Cho} \snu
\author{Alexei V. Filippenko} \berkeley \miller
\author{John C. Horst} \sandiego
\author{Dragana Ili\'c} \belgrade
\author{Michael. D. Joner} \byu
\author{Daeun Kang} \snu
\author{Wonseok Kang} \nysc
\author{Shai Kaspi} \telaviv
\author{Taewoo Kim} \nysc \affil{Department of Astronomy and Space Science, Chungbuk National University, Cheongju 28644, Korea}
\author{Andjelka B. Kova\v{c}evi\'c} \belgrade
\author{Sahana Kumar} \berkeley \fsu
\author{Huynh Anh N. Le} \snu \hefei
\author{A. E. Nadzhip} \sai
\author{Francisco Pozo Nu\~nez} \affil{Haifa Research Center for Theoretical Physics and Astrophysics, University of Haifa, Haifa, Israel}
\author{V. G. Metlov} \sai
\author{V. L. Oknyansky} \sai
\author{Songyoun Park} \snu
\author{Luka \v{C}. Popovi\'c} \aob
\author{Suvendu Rakshit} \affil{Indian Institute of Astrophysics, Block II, Koramangala, Bangalore-560034, India} \snu \affil{Finnish Centre for Astronomy with ESO (FINCA), University of Turku, Quantum, Vesilinnantie 5, FI-20014 University of Turku, Finland}
\author{Malte Schramm} \affil{National Astronomical Observatory of Japan, Mitaka, Tokyo 181-8588, Japan}
\author{N. I. Shatsky} \sai
\author{Michelle Spencer} \byu
\author{Eon-Chang Sung} \kasi
\author{Hyun-il Sung} \kasi
\author{A. M. Tatarnikov} \sai
\author{Oliver Vince} \aob

\correspondingauthor{Jong-Hak Woo}\email{woo@astro.snu.ac.kr}

\begin{abstract}
We present a variability study of the lowest-luminosity Seyfert 1 nucleus of the galaxy NGC 4395 based on photometric monitoring campaigns in 2017 and 2018. Using 22 ground-based and space telescopes, we monitored NGC 4395 with a $\sim$5 minute cadence during a period of 10 days and obtained light curves in the ultraviolet (UV), $V$, $J$, $H$, and $K/Ks$ bands as well as narrow-band \Ha. The root-mean-square (RMS) variability is $\sim 0.13$ mag in the \swift{}-{\it UVM2} and $V$ filter light curves, decreasing down to $\sim 0.01$\,mag in the $K$ filter. After correcting for the continuum contribution to the \ha\ narrow-band, we measured the time lag of the \Ha\ emission line with respect to the $V$-band continuum as \valerrud{55}{27}{31} to \valerrud{122}{33}{67} min in 2017 and \valerrud{49}{15}{14} to \valerrud{83}{13}{14} min in 2018, depending on assumptions about the continuum variability amplitude in the \Ha\ narrow-band. We obtained no reliable measurements for the continuum-to-continuum lag between UV and $V$ bands and among near-IR bands, owing to the large flux uncertainty of UV observations and the limited time baseline. We determined the active galactic nucleus (AGN) monochromatic luminosity at 5100\,\AA, $\lambda L_\lambda = \left(5.75\pm0.40\right)\times 10^{39}\,\mathrm{erg\,s^{-1}}$, after subtracting the contribution of the nuclear star cluster. While the optical luminosity of NGC 4395 is two orders of magnitude lower than that of other reverberation-mapped AGNs, NGC 4395 follows the size-luminosity relation, albeit with an offset of 0.48\,dex ($\geq 2.5\sigma$) from the previous best-fit relation of \citet{Bentz+13}.
\end{abstract}

\section{Introduction}
The origin of supermassive black holes is a subject of intensive research, including theoretical investigations on the black hole seeds in various mass scales \citep[e.g.,][]{Volonteri+03, Volonteri+08, Barai&deGouveiaDalPino19}. Direct observational evidence of the fossil record of either light or heavy seeds at high redshift is currently unavailable. The grown stage of supermassive black holes and their occupation fraction have been investigated to unveil the growth history of supermassive black holes \citep[e.g.,][]{Miller+15, Gallo&Sesana19, Reines+13}.

While the mass of the dynamically confirmed supermassive black holes is typically larger than a million solar masses in the present-day universe, it is unclear whether a population of intermediate-mass black holes exists at the center of less-massive galaxies \citep[e.g.,][]{Greene12}. It is observationally challenging to reveal the presence of intermediate-mass black holes, as the dynamical measurements suffer from a limited spatial resolution even with the best available observational facilities. To probe the sphere of influence of an intermediate-mass black hole, $\Mbh<<10^6$ \msun\ (i.e., $R_{\mathrm{inf}}= G\Mbh / \sigma_{\ast}^2$, where $\sigma_{\ast}$ is the stellar velocity dispersion), typically a resolution better than 1\,pc is required. For example, \cite{Nguyen+17} reported an upper limit of $1.5 \times 10^5$ \msun\ of the central black hole in NGC 404 based on dynamical measurements with $\sim 1$\,pc resolution, demonstrating the challenge of finding dynamical evidence of intermediate-mass black holes. There have been dynamical mass measurements or upper limits of intermediate-mass black holes for only a small number of local galaxies \citep{denBrok+15, Nguyen+18, Nguyen+19, Greene+19}. While there are also various reports on intermediate-mass black holes, such as in ultraluminous X-ray sources \citep[e.g.,][]{Mezcua+18}, the origin of supermassive black holes is more closely related to the black holes at the galaxy centers, which are believed to be connected with their host galaxies in the growth history over the Hubble time \citep{Kormendy&Ho13}. 

The nearby galaxy NGC 4395 at a distance of 4.4 Mpc \cite{denBrok+15} is a unique testbed for studying intermediate-mass black holes. As a Seyfert 1 galaxy, NGC 4395 hosts an active galactic nucleus (AGN) with extremely low luminosity; the bolometric luminosity is typically reported to be below $10^{41}\,\mathrm{erg\,s^{-1}}$ \citep[e.g.,][]{Filippenko&Sargent89, Filippenko+93, Lira+99, Moran+99, Filippenko&Ho03}. The host galaxy is classified as a dwarf galaxy with a stellar mass $10^{9}$ \msun\ \citep{Filippenko&Sargent89, Reines+13}, and there is no clear sign of a bulge while a bar-like central structure is identified \citep{denBrok+15}. Various studies determined the mass of the central black hole in NGC 4395. For example, \citet{Filippenko&Ho03} estimated $\Mbh=1.3\times10^4$ \msun\ using the broad-line region (BLR) size-luminosity relation from \cite{Kaspi+00}. \cite{Edri+12} also reported $\Mbh=\left(4.9\pm 2.6\right)\times 10^4$ \msun\ based on the reverberation mapping analysis of Balmer broad emission lines using broad-band photometry \citep[see also][]{Desroches+06}. On the other hand, \citet{Peterson+05} reported a higher mass, $\Mbh=\left(3.6\pm 1.1\right)\times 10^5$ \msun, using \civ{} line reverberation mapping.

Recently, \citet{Woo+19} reported the mass of the central black hole in NGC 4395 as $\Mbh=\verrud{9100}{1500}{1600}$ \msun\ based on a reverberation mapping analysis, using narrow-band photometry. The reported time lag of the \ha\ emission, 83 min, is longer than that of the \civ\ emission line, 48--66 min \citep{Peterson+05}, suggesting consistency with the stratification of the BLR, which leads to a factor of $\sim 2$ longer lag for \Ha\ than \civ. In contrast, \citet{Woo+19} measured the line dispersion velocity of \Ha\ to be $\sigma=426\pm1$\,km\,s$^{-1}$, while the line dispersion velocity of \civ\ was reported as $\sigma \approx 2900$\,km\,s$^{-1}$ by \citet{Peterson+05}. Therefore, the main discrepancy of the black hole mass between \Ha-based and \civ-based reverberation mapping results is from the line-width measurements. There have been various studies investigating the systematic difference between \Hb- and \civ-based black hole masses, and in general \civ-based mass suffers more uncertainties \citep[e.g.,][]{Park+13, Denney+13}.

While NGC 4395 presents an intermediate-mass black hole and low luminosity, the Eddington ratio of NGC 4395 is $\sim 5$\% \citep{Woo+19}, which is comparable to that of other reverberation-mapped AGNs. Therefore, it provides a useful testbed for investigating the effect of luminosity vs. accretion rate on AGN properties -- e.g., BLR stratification, nonvirial motions, X-ray spectral energy distribution (SED), etc.

Having the lowest-luminosity Seyfert 1 nucleus known today, NGC 4395 also provides an interesting opportunity to investigate photoionization and the size-luminosity relation at the low-luminosity regime. The size-luminosity relation has been studied with AGNs having optical luminosity at 5100\,\AA\ larger than 10$^{42}$ \ergs\ \citep{Bentz+13}, while it is yet to be probed whether the photoionization assumption is valid at extremely low luminosity. 
  
In this study, we present variability analysis using data from our NGC 4395 monitoring campaign in 2017 and 2018, which include ultraviolet (UV), optical, and near-infrared (IR) photometry. We investigate the effect of the continuum on the narrow-band light curves for constraining the validity of the lag measurements. Also, we investigate the BLR size-luminosity relation \citep{Bentz+13} at the extreme low-luminosity end by combining NGC 4395 with previous reverberation results of AGNs with measured supermassive black holes. In Section 2, we describe the observations and data-reduction processes, and we present the data analysis in Section 3. We compare our results with those previously published and discuss the BLR radius-luminosity relation in Section 4. Our results are summarized in Section 5. We assume the distance to NGC 4395 to be 4.4 Mpc throughout this paper, as adopted by \cite{denBrok+15}.

\section{Observation and Data Reduction}

In 2017 and 2018 we performed intensive monitoring campaigns using 22 ground-based and one space telescope in order to obtain well-sampled light curves of the AGN continuum flux and \Ha\ emission-line flux over a few days timescale. The time lag of the \Ha\ emission line with respect to the V-band continuum is roughly estimated to be 1--4\,hr based on the monochromatic luminosity at 5100\,\AA\ and the \Hb\ size-luminosity relation \citep{Bentz+13}. Thus, a period of approximately one day (24\,hr) of observations is required to obtain a sufficiently long temporal baseline to track the time lag between the V-band and \Ha\ light curves properly. Therefore, we combined multiple telescopes at various longitudes to fill in the daytime gap at each telescope. The details of the participating observatories along with their telescopes and instruments are summarized in Table~\ref{table:telescopes}. 

The campaign was carried out from 28 April to 5 May 2017 (all dates are presented in UT), and from 6 to 9 April 2018, during which we used various filters covering the UV to the near-IR continuum. 

\renewcommand{\arraystretch}{1.25}
\begin{deluxetable*}{lrccc}[t]
	\tablewidth{0.99\textwidth}
	\tablecolumns{5} 
	\tablecaption{Observing Facilities Participating in the Campaign\label{table:telescopes} }
	\tablehead
	{
		\colhead{Observatory Name}&\colhead{Longitude}&\colhead{Aperture}&\colhead{Detector}&\colhead{Filters}
	}
	\startdata
	Bohyunsan Optical Astronomy Observatory (BOAO) & $128^{\circ}58'$E & 1.8m & e2v CCD231-84 & V, \ha 
	\\
	Mt. Laguna Observatory (MLO) & $116^{\circ}25'$W & 1m & e2V 42-40 2k & B, V 
	\\
	MDM Observatory (MDM) & $111^{\circ}37'$W & 1.3m & Templeton & B, V, \ha 
	\\
	&  & 2.4m & MDM4K & V, \ha
	\\
	Mt. Lemmon Optical Astronomy Observatory (LOAO) & $110^{\circ}47'$W & 1.0m & e2v CCD 231-84 & B, V 
	\\
	West Mountain Observatory (WMO) & $111^{\circ}50'$W & 0.9m & FLI-PL3041-UV & B, V 
	\\
	Caucasus Mountain Observatory (CMO) & $42^{\circ}40'$E& 0.6m & Aspen CG42 & B, V 
	\\
	& & 2.5m & HAWAII 2-RG & J,H,K
	\\
	Astronomical Station Vidojevica (ASV) & $21^{\circ}33'$E & 1.4m & Apogee U42 & B, V 
	\\
	Sobaeksan Optical Astronomy Observatory (SOAO) & $128^{\circ}27'$E & 0.61m & e2v CCD42-40 & B, V 
	\\
	Vainu Bappu Observatory (VBO) & $78^{\circ}49'$E & 1.3m & Apogee Aspen CG42 & B, V 
	\\
	Higashihiroshima Astronomical Observatry (Hiroshima) & $132^{\circ}47'$E & 1.5m & HOWPol & B, V, R 
	\\
	LCOGT - McDonald (McDonald) & $104^{\circ}01'$W & 1m & Sinistro & V 
	\\
	LCOGT - Haleakala (Haleakala) & $156^{\circ}15'$W & 0.4m & SBIG 6303 & V 
	\\
	Lick Observatory (Nickel) & $121^{\circ}39'$W & 1m & Direct Camera CCD-2 & B, V
	\\
	Wise Observatory (Wise) & $34^{\circ}46'$E & 1m & STX-16803 & B, V, \ha 
	\\
	&  & 0.7m & FLI-PL16801 & B, V 
	\\
	Cerro Tololo Inter-American Observatory (CTIO) & $70^{\circ}48'$W & 0.6m & SBIG STL-11000M & V, \ha
	\\
	Okayama Astrophysical Observatory (OAO) & $133^{\circ}36'$E & 0.91m & OAO/WFC & $\mathrm{K_S}$
	\\
	Deokheung Optical Astronomy Observatory (DOAO) & $127^{\circ}27'$E & 1m & PL-16803\tablenotemark{a} & B, V
	\\
	& & & SOPHIA-2048B\tablenotemark{b} & B, V
	\\
	Dark Sky Observatory (DSO) & $81^{\circ}25'$W & 0.36m & Apogee Alta U47 & V
	\\
	\swift\ Ultraviolet/Optical Telescope (UVOT) & \emph{satellite} & 0.3m & Intensified CCD & UVM2 \\
	Gemini Observatory - North & $155^{\circ}28'$W & 8.1m & GMOS-N & g, Spectroscopy 
	\\
	\enddata
	\tablenotetext{a}{Used in 2017}
	\tablenotetext{b}{Used in 2018}
\end{deluxetable*}

\subsection{Optical Observations}\label{ss:photometry}

In the optical range, we mainly used the V-band and narrow \Ha-band filters, while the B and R-band filters were occasionally used for flux calibration. As the time lag of the \Ha\ emission line with respect to the optical continuum is expected to be $\lesssim 1\hr$, a relatively short time cadence was required. For the continuum monitoring with the V-band filter, we mainly used 1-m-class telescopes, while for the narrow \Ha-band monitoring we used $\sim 2$-m-class telescopes to increase the signal-to-noise ratio (S/N) per exposure.

\begin{figure*}
\center
\includegraphics[width=0.45\textwidth]{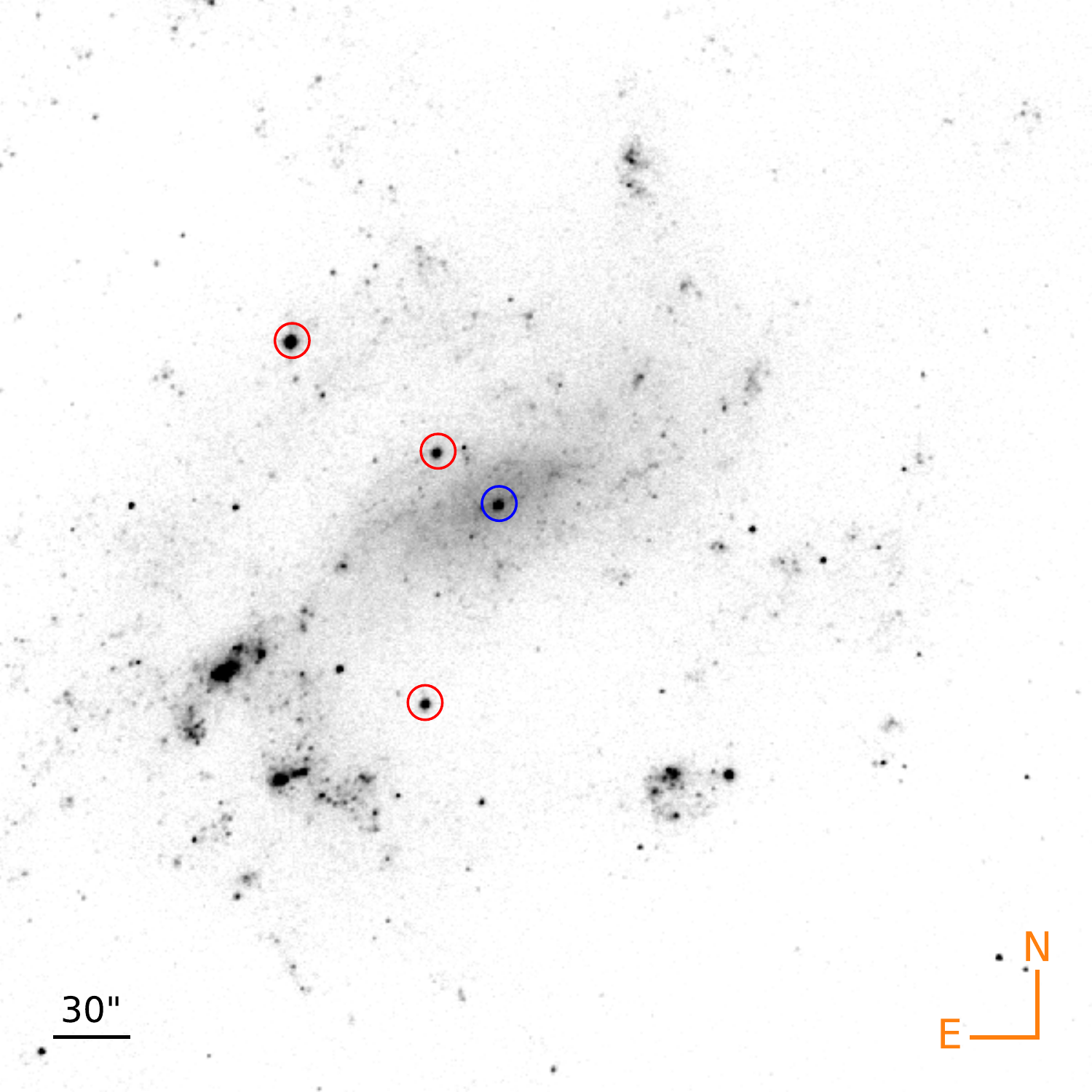}
\includegraphics[width=0.45\textwidth]{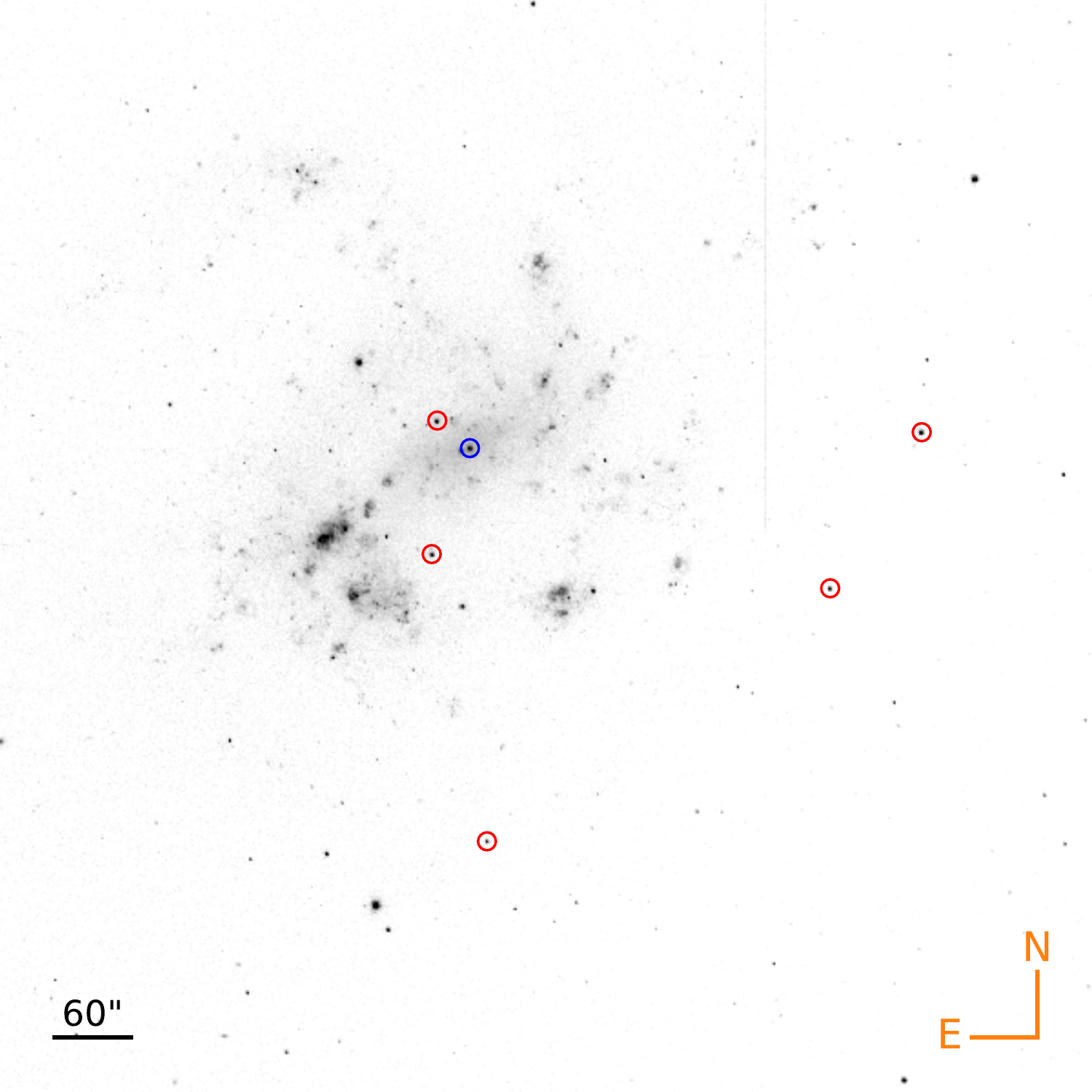}
\caption{\emph{Left}: An example of V-band images of NGC 4395 with 300\,s exposure, obtained at the MDM 1.3 m telescope. Three comparison stars are marked with red circles, while the target AGN is denoted with a blue circle. The field of view is $7.6' \times 7.6'$ after cropping, and the radius of the circle corresponds to the aperture size of $4''$ for differential photometry. 
\emph{Right}: An example of the \ha\ narrow-band  images of NGC 4395 with a field of view $14.3'\times14.3'$ after cropping, obtained at the MDM 2.4m telescope. The blue circle denotes the AGN, whereas red circles mark 5 comparison stars. The field of view is shown after trimming the shadow of the guide probe and the vignetted region. The radius of the circle corresponds to the aperture size of $7''$ for differential photometry.	
}
\label{fig:FOV} 
\end{figure*}

Examples of the images of NGC 4395 taken from different telescopes and bands are shown in Figure~\ref{fig:FOV}. To ensure better than 1--2\% flux measurement errors, we determined the optimal exposure time for each telescope based on the imaging data, which were obtained before the start of the monitoring campaign. For example, we used 180\,s exposure time for the V-band imaging with the MDM 2.4m telescope, while in general we used 300\,s exposure time for both the V-band and \Ha. Thus, we maintained $\sim 5$\,min time resolution at each telescope.

\begin{figure}
	\center
	\includegraphics[width=0.45\textwidth]{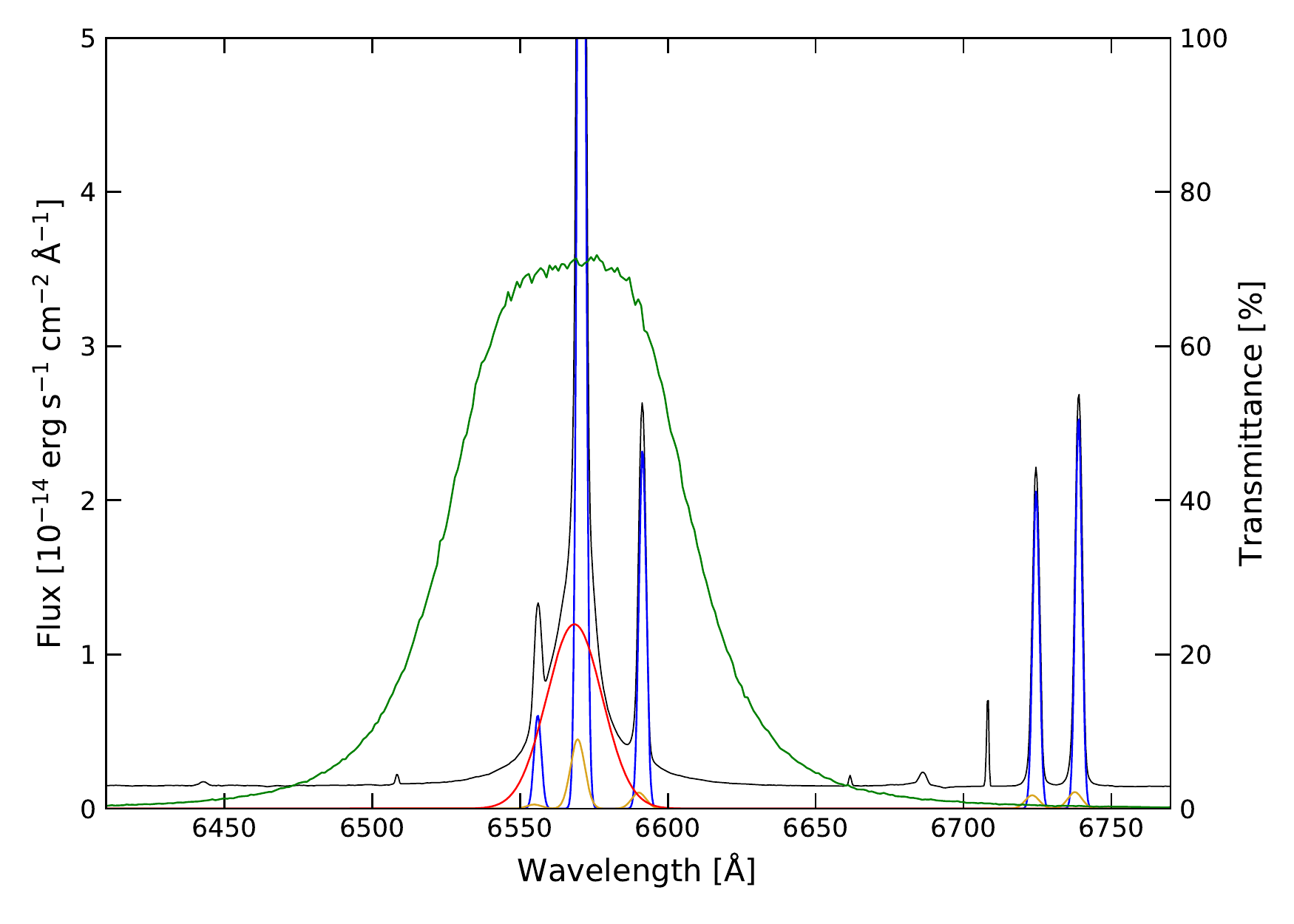}
	\caption{The response function of the KP1468 filter available at the MDM 1.3m telescope (green line). The spectrum of NGC 4395 obtained with the Gemini GMOS is compared with the response function. Note that the emission lines were decomposed into broad \Ha\ component (red), and a wing (yellow) and core component (blue) of the narrow lines (i.e., \Ha, \NII, \SII). }
	\label{fig:contfrac} 
\end{figure}

For the narrow \Ha-band observations, we used the MDM 1.3m, BOAO 1.8m, and MDM 2.4m telescopes in 2017. The weather at the MDM observatory was relatively good, while the data from BOAO suffered large uncertainties due to bad weather and low sensitivity; hence, these data were not used in the cross-correlation analysis. In 2018, we used the MDM 2.4m telescope for the narrow \Ha-band observations. Since we only used the \Ha-band data from the MDM 2.4m telescope for the time-lag analysis, we present the response function of the narrow \Ha-filter in Figure~\ref{fig:contfrac}, which covers a spectral range of 6470--6560\,\AA, including the broad \Ha\ emission line along with the narrow \NII\ and \Ha\ lines. Note that while the flux from emission lines is dominant in the narrow \Ha-band, there is a significant contribution from the continuum, which has to be taken into account to obtain a reliable lag for the \Ha\ emission line (see \ref{ss:contcorr}). 

Standard data reduction was performed including bias subtraction and flat-fielding using \textit{IRAF}\footnote{\textit{IRAF} is distributed by the National Optical Astronomy Observatory, which is operated by the Association of Universities for Research in Astronomy (AURA), Inc., under a cooperative agreement with the U.S. National Science Foundation (NSF).} procedures, and cosmic ray rejection using the L.A.Cosmic algorithm \citep{vanDokkum+01}. If necessary, 2--4 consecutive exposures were combined to construct a single-epoch image to decrease the photometric uncertainty to $<5$\%, while the time resolution between epochs was kept at a maximum of 10 min. After that, data quality was assessed based on visual inspection, and any epoch with quality issues (e.g., failed tracking or performance trouble reported in the observing log) was rejected from further photometric analysis. 

\begin{figure}
	\center
	\includegraphics[width=0.49\textwidth, height=14cm]{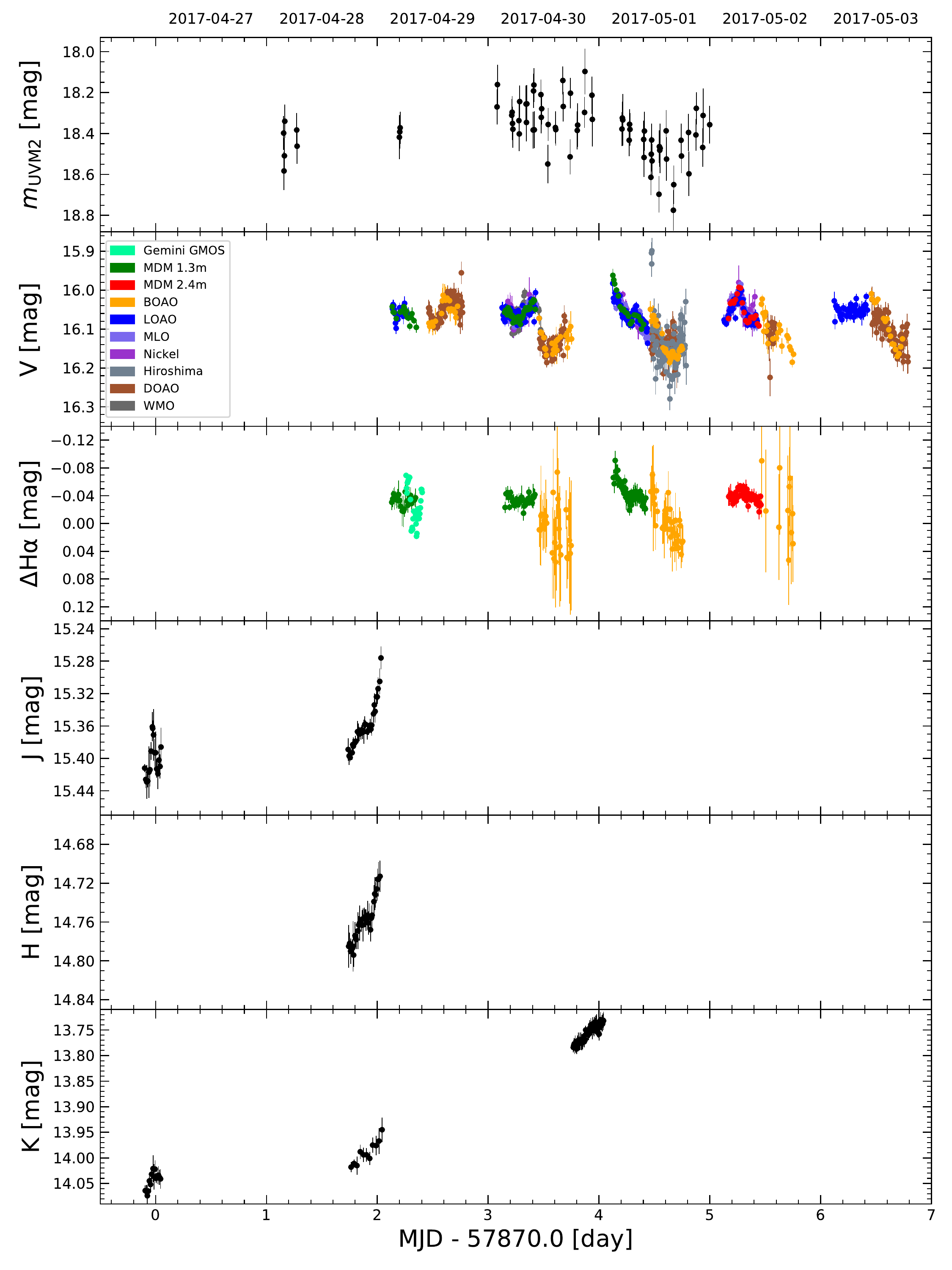} 
	\caption{Light curves in the UV, optical (V and \ha{}), and near-IR bands (J, H, and K) obtained in 2017. For the optical data, we only present the light curves from 10 telescopes with good weather conditions during the observations and the mean error in V-band photometry $< 0.03$\,mag. 
	The \ha{} panel shows light curves obtained from an \ha\ narrow-band filter, as well as a light curve obtained from GMOS spectral modeling. 
	All uncertainties shown here are 1$\sigma$. \label{fig:lc2017}} 
\end{figure}

\begin{figure}
	\center
	\includegraphics[width=0.49\textwidth]{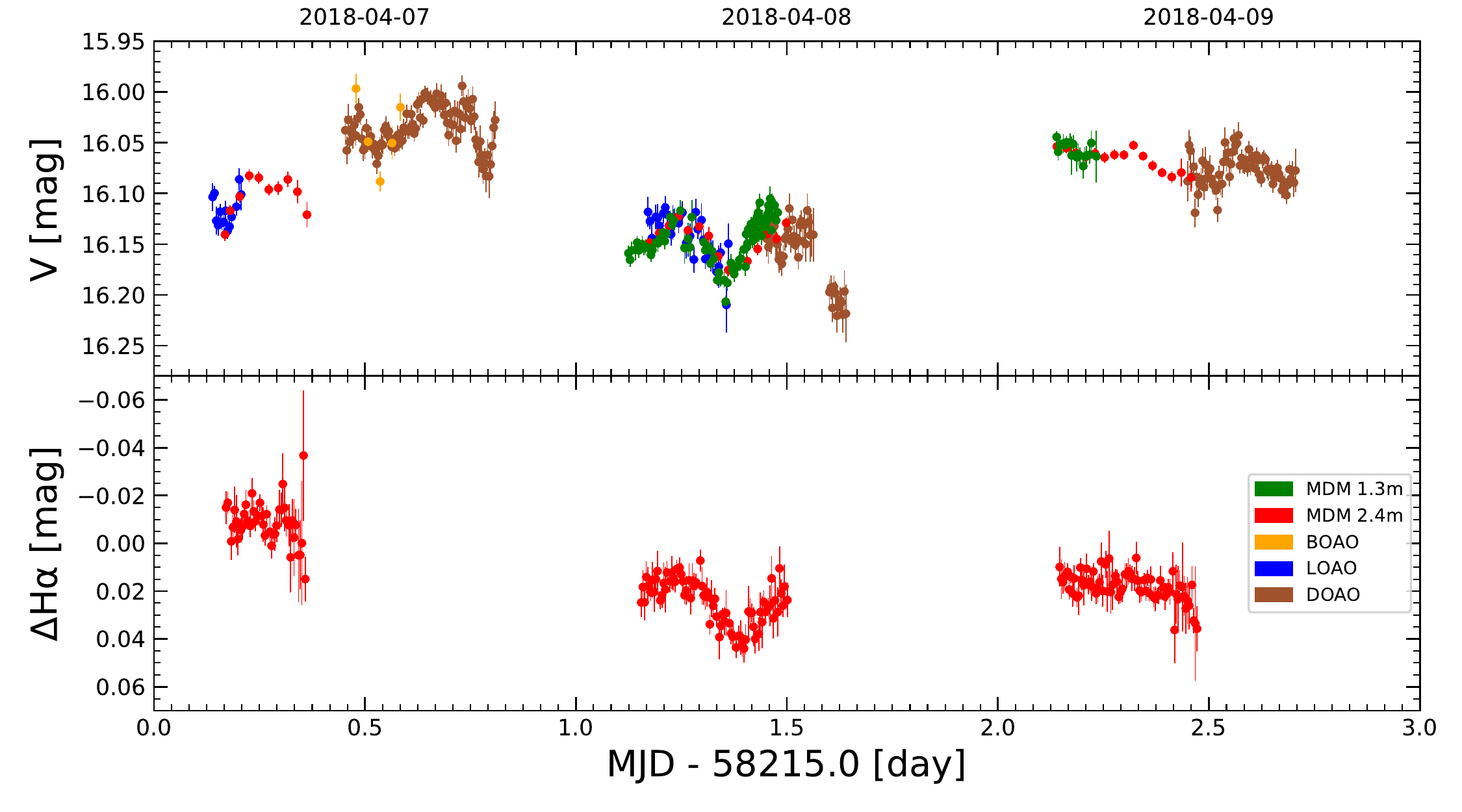}
	\caption{V (\emph{top}) and \ha\ (\emph{bottom}) light curves obtained in 2018. We only show the light curves from 5 telescopes with good weather conditions during the observations and average error in V-band light curve $< 0.03$\,mag. All uncertainties shown here are 1$\sigma$. \label{fig:lc2018}} 
\end{figure}

\subsection{UV Observations}\label{ss:uv}

We observed NGC 4395 using the \swift\ UVOT to monitor the variability of the UV continuum and to investigate the time lag between continuum bands. The UVOT data were taken from 2017-04-28 to 2017-05-02 using the UVM2 filter, which is centered at 2231\,\AA. Note that one orbital period of \swift\ is 96 min, of which NGC 4395 was visible for $\sim 2000$\,s. We performed the data reduction with HEASoft \citep{FTOOLS}, including background subtraction, correction for anomalous zero exposures, and correction for the degradation of the UVOT sensitivity.

\subsection{Near-Infrared Observations}\label{ss:ir}
We monitored NGC 4395 using J, H, and K/Ks filters at the Caucasus Mountain Observatory (CMO) 2.5m telescope \citep[the ASTRONIRCAM instrument,][]{Nadjip+17} and Okayama Astrophysical Observatory (OAO) 0.91m telescope \citep[OAO/WFC, ][]{Yanagisawa+19}. At the CMO 2.5m, we used the K band filter of the Mauna Kea Observatory (MKO) system for 3 nights in 2017. In addition, J and H filters were occasionally used during the monitoring. In the case of OAO observations, we used the Ks filter, but the image quality was too poor to perform further analysis.

\subsection{Optical Spectroscopic Observations}
In addition to the photometric observations, we performed spectroscopic observations for $\sim$3.5\,hr, using the Gemini Multi-Object Spectrograph (GMOS) on the Gemini North telescope on 29 April 2017. Initially, we planned to observe NGC 4395 for two consecutive nights in 2017 (GN-2017A-Q2, PI: Woo), and 3 nights in 2018 (GN-2018-Q102, PI: Woo), but all scheduled nights were lost due to snowstorms. Nevertheless, we were able to monitor the target for $\sim$3.5\,hr under varying cloudy conditions on 29 April 2017. 

We used a long slit with a $0.75''$ width and the R831 grating, obtaining a spectral resolution of $R=2931$, good enough to resolve the broad and narrow components of \Ha, and the wing and core components of the narrow lines, \NII\ and \SII\ as presented by \citet{Woo+19}. The instrument setup covered the spectral range 4606--6954\,\AA\ with a 0.374\,\AA\ pix${-1}$ scale. We set the position angle to be 50.2 degrees East from North. We used 2-pixel binning along the spatial direction, resulting in a scale of 0.16\arcsec\ pix$^{-1}$. Each exposure was 300\,s long and we obtained a high-quality spectrum every 6 min, including 1 min overhead per exposure. A total of 36 exposures was obtained during the 3.5\,hr run, but three epochs were discarded owing to strong cosmic rays that hit the \ha\ line. Additionally, we observed G191-B2B for flux-calibration purpose \citep{Massay+88, Massay&Gronwall90, Bohlin+95}.

We performed standard data reduction with the Gemini IRAF package, including bias subtraction, flat fielding, wavelength calibration, and flux calibration. Cosmic rays were rejected using  the L.A.Cosmic routine \citep{vanDokkum+01}. From each exposure, we extracted a one-dimensional spectrum using an aperture size that was three times the seeing full width at half-maximum intensity (FWHM), in order to compensate for varying seeing during the observing run. Out of the 33 epochs, two consecutive spectra showed relatively low S/N, so we averaged these two epochs. Thus, we finalized a total of 32 spectra.

To obtain the flux of the broad \Ha\ emission line, we performed decomposition analysis as outlined by \citet{Woo+19}. In brief, we modeled the \SII\ $\lambda\lambda 6717$, 6731 doublet using two Gaussian components for each line since both \SII\ lines exhibit a broad wing component and a narrow core component. For the continuum subtraction, we adopted a straight line using the continuum flux around the 6660--6700\,\AA\ and 6760--6800\,\AA{\ ranges. By assuming that the line profile and flux of \SII\ were constant during the night, we performed a calibration for the flux, spectral resolution, and wavelength shift, so that the \SII\ $\lambda\lambda 6717$, 6731 model profile of each epoch remains constant. Also, we constructed mean and root-mean-square (RMS) spectra using the calibrated spectra.

We then modeled the \ha\ and \NII\ region by assuming that the narrow \ha\ line and the \NII\ $\lambda\lambda 6548$, 6583 doublet have the same profile as that of \SII\ $\lambda\lambda 6717$, 6731. For each epoch, we modeled the broad \ha\ line with a single Gaussian profile, and the narrow lines (\ha\ and \NII) with the same profile as obtained for \SII. In addition, we used a linear  continuum. We simultaneously fitted emission lines and continuum in the spectral range 6450--6670\,\AA. We obtained the light curve of the broad \ha\ line flux as shown in the second panel in Figure~\ref*{fig:lc2017}. Note that the $\sim$3.5\,hr baseline is too short to reliably measure the time lag between the continuum and the broad emission line (see \ref{ss:lag}).

\section{Analysis}

\subsection{Differential Photometry}\label{ss:diffphot}

To measure the flux variability of the AGN continuum and \Ha\ emission line, we first performed aperture photometry, using the \texttt{photutils} package \citep{photutils_v0.4}. Since the majority of images showed seeing variations during the night, we matched the point-spread function (PSF) before performing aperture photometry. The PSF was constructed based on isolated, bright, and unsaturated stars in each image, which were subsequently convolved so that all images have the same matched PSF, which was obtained from the worst-seeing image of the night, typically 1.5--$4''$. Using the PSF-matched images, we then performed aperture photometry for the target AGN and comparison stars in the field of view. A global background image was constructed using the \texttt{SExtractorBackground} estimator of the \texttt{photutils} package, which was subtracted from each image.

In addition, we determined the residual background for individual sources using annuli with an outer radius of 3--5 times the seeing FWHM and an inner radius of 2--3 times the seeing FWHM. The calculated residual background value was then subtracted from the median value measured within the aperture. We note that the residual background flux is insignificant and the additional background subtraction did not change the flux of most of the comparison stars. In contrast, this process was required for NGC 4395, since the host-galaxy contribution at the galactic center was significant. Consequently, the additional subtraction decreased the AGN flux. 

We determined the aperture size to include more than 99\% of the point-source flux and performed differential photometry for a number of nearby comparison stars. Depending on the field of view of each camera at each telescope, we selected various numbers of bright stars (3--8 stars for each set of observations) which were nonvarying and unsaturated with photometric uncertainty $<2$\%. To identify variable stars, we used the table from \cite{Thim+04}.

In Figure 1, we present an example of the V-band and \Ha-band images along with the selected comparison stars. We calculated the difference between the instrumental magnitude and the known magnitude of each comparison star, and adopted the mean difference as the relative normalization value ($\delta$V) for each epoch. We assumed the standard deviation of the $\delta$V from individual comparison stars as a systematic error of the normalization, which was added to the uncertainties of the instrumental magnitude and the background flux for calculating a total uncertainty. 

Based on the aperture photometry and calibration, we constructed light curves by combining measurements from different telescopes. In this process, we intercalibrated the light curves in order to avoid a systematic offset between light curves obtained from different telescopes due to differences in the response function of the filters, detector efficiency, etc. We matched each light curve with a reference light curve by adding a linear shift in magnitude. In other words, the mean magnitude within the overlapped time interval in a light curve was forced to be the same as that of the reference light curve.

\subsubsection{Detailed Intercalibration Information}\label{sss:intercal}
For the light curve of 2017-04-30, WMO, MLO, and BOAO data were intercalibrated with each other as the light curve of WMO overlaps with the light curve from the other two telescopes. We applied the same correction to those of the MLO and BOAO on other nights. The light curves of MDM 1.3m on 2017-04-30 and 2017-05-01 were calibrated with respect to those of MLO on respective nights, and the average correction shift for the MDM 1.3m was applied to its light curve of 2017-04-28. Finally, any light curves that overlapped with BOAO were calibrated with respect to BOAO, while any light curves that overlapped with MLO or MDM 1.3m were calibrated to them. 

For the 2018 data, we calibrated light curves of DOAO with respect to those of MDM 2.4m on 2018-04-08 and 2018-04-09, and the average correction factor shift was applied to the light curve of DOAO 2018-04-07. Other light curves were calibrated with respect to either MDM 2.4m or DOAO, depending on which light curve they overlap the most. Finally, we calculated V-band zero points using the bright nearby star 2MASS J12255090+3333100, whose V-band magnitude was determined as $\mathrm{V_{*}}=16.9$ by converting the \emph{u'g'r'i'z'} magnitudes from the Sloan Digital Sky Survey Data Release 12 (SDSS DR12) based on the equations of \cite{Jester+05}. Comparing with the instrumental magnitude of the comparison star, we obtained the normalization and applied it to the V-band light curve obtained with the MDM 2.4m telescope on 2017-05-02 and 2018-04-08, and all other intercalibrated light curves were adjusted accordingly. 

Photometric \ha\ light curves were not intercalibrated in the same manner as the V-band was calibrated since their light curves did not overlap with each other; for the same reason, intercalibration did not affect the relative photometry of \ha . We only shifted the light curve of the \ha\ broad component from Gemini GMOS spectroscopic observation to match that of the MDM 1.3m. 

\subsubsection{Light Curves}\label{sss:LC}
In Table~\ref*{table:lc}, Figure~\ref*{fig:lc2017}, and Figure~\ref*{fig:lc2018}, we present the calibrated light curves of the V-band and \ha-band, after excluding the data obtained during bad weather since the error bars are too large to provide any meaningful measurements. Overall, the two curves show similar trends, demonstrating correlated variability. However, most segments of the V-band light curve are not suitable for reverberation-mapping analysis owing to (1) the lack of corresponding \Ha\ observations, (2) the weak variability in the \Ha-band light curve, and (3) the large uncertainty of the \Ha\ photometry. Among the monitoring data obtained in 2017 and 2018, we identified only two nights, 2017-05-02 and 2018-04-08, from which we were able to achieve reliable time-lag measurements. We will focus on these two nights for further reverberation-mapping analysis in Section \ref{ss:contvar}.

\renewcommand{\arraystretch}{1.25}
\begin{deluxetable}{ccccl}[t]
	\tablewidth{0.95\textwidth}
	\tablecolumns{5} 
	\tablecaption{Photometry Data\label{table:lc}}
	\tablehead
	{
		\colhead{MJD-50000.0}&\colhead{Band}&\colhead{Magnitude}&\colhead{Uncertainty}&\colhead{Notes}
		\\
		\colhead{(1)}    &\colhead{(2)}       &\colhead{(3)}       &\colhead{(4)}&\colhead{(5)}   
	}
	\startdata 
	7869.90099 &	J &	15.412 &	0.005 & \\
	7869.90569 &	K &	14.064 &	0.010 & \\
	7869.91038 &	J &	15.426 &	0.011 & \\
	7869.91508 &	K &	14.065 &	0.012 & \\
	7869.91931 &	J &	15.429 &	0.021 & 
	\enddata
	\tablecomments{Columns are (1) Modified Julian date, (2) filter, (3) magnitude, (4) 1$\sigma$ uncertainty in magnitude, and (5) note. \\(This table is available in its entirety in machine-readable form.)}
\end{deluxetable}

For the data obtained with the \swift\ UVM2 filter, we binned event files into 300\,s exposures and measured the count rate from circular apertures of $3''$ radii in order to maximize the S/N. Photon count rates, $\dot{\gamma}$, were then corrected for the large-scale sensitivity gradient and converted into AB magnitudes as $m_\mathrm{UVM2} = -2.5 \log_{10} \dot{\gamma} + 18.54$ \citep{Breeveld+11}. Finally, aperture correction to the standard UVOT aperture of $5''$ was applied using on-field stars in each temporal bin in order to convert aperture magnitude into PSF magnitude. 

For the J, H, and K images, AGN magnitudes were obtained for each exposure by performing photometry with circular apertures of $2.2''$ diameter, with respect to the comparison star 2MASS J12255090+3333100, whose magnitude in each band is $\mathrm{J}=14.362$, $\mathrm{H}=13.939$, and $\mathrm{K}=13.786$ when converted to the MKO system. 

\subsection{Variability}
We quantified the variability of NGC 4395 using the light curves presented in Figures \ref*{fig:lc2017} and \ref*{fig:lc2018}. We calculated the RMS variability in magnitude ($\sigma_{m}$), the ratio between the maximum flux and minimum flux ($R_\mathrm{max}$), and the fractional variability ($\fvar$), using the 2017 and 2018 light curves. The fractional variability $\fvar$ is defined as 
\begin{align}
\fvar = \frac{1}{\langle{f}\rangle}\sqrt{\langle f^2 \rangle - \langle f \rangle^2   -\left\langle {\epsilon_f}^2 \right\rangle}
\label{eq:fracvar}
\end{align}\citep{Vaughan+03},
where $f$ is the flux at each epoch, $\langle f \rangle$ is the mean flux, and $\epsilon_f$ is the flux uncertainty. The error of the fractional variability is given by
\begin{align}
\varepsilon(\fvar) = \sqrt{
	\left(\sqrt{\frac{1}{2N}}\frac{\left\langle {\epsilon_f}^2 \right\rangle}{\left\langle f \right\rangle^2 \fvar}\right)^2
	+
	\left(\sqrt{\frac{\left\langle {\epsilon_f}^2 \right\rangle}{N}\frac{1}{\left\langle {f} \right\rangle}}\right)^2,
}\label{eq:fracvarerr}
\end{align}
where $N$ is the number of epochs.

Table~\ref{table:varstat} summarizes the variability measurements. Using the light curves obtained in 2017, we find RMS variability from 0.02 to 0.13 mag in the continuum, and the 2018 data show a similar range. $R_\mathrm{max}$ of the continuum band ranges from 1.08 to 1.87 in 2017 and similar values in 2018. Accounting for the measurement errors in the light curves from 2017, we find that the fractional RMS variability ranges from 1\% to 8\%, showing a decreasing trend with increasing continuum wavelength. The exception is the UVM2 band, which has a similar fractional variability compared to the V band, due to the larger uncertainty of the UV photometry. In general, we find similar trends in the light curves obtained in 2018. 

To compare the variability with the consistent length of the time baseline, we calculated the variability statistics of the near-IR bands using the light curves obtained on 2017-04-28. RMS deviation, $R_\mathrm{max}$ variability, and fractional variability show a clear decreasing trend with increasing wavelength. 

We also used the data from 2017-05-02 and 2018-04-08 to measure the variability in the V band and the \ha\ narrow-band for comparison. We obtained RMS variability of 0.01 mag, $R_{\rm max}$ of 1.1, and fractional RMS variability of 1--2\%. Note that since the narrow band contains nonvariable narrow lines (\NII\ and narrow \Ha\ emission) which account for 49\% of the total flux observed in the \Ha\ narrow-band (see \ref{ss:contcorr}), the actual variability amplitude is a factor of 3 higher than these measurements. We also measured the variability of the entire \ha\ light curves of 2017 and 2018, and obtained $\sigma_{m,\,\mathrm{H\alpha}}=0.012$, $R_\mathrm{max,\, H\alpha}=1.07$, and $\fvar_{,\, \mathrm{H\alpha}}=0.006$ in 2017 and $\sigma_{m,\,\mathrm{H\alpha}}=0.015$, $R_\mathrm{max,\, H\alpha}=1.08$, and $\fvar_{,\, \mathrm{H\alpha}}=0.012$ in 2018, which are broadly consistent with single-night values with continuum correction. 

\renewcommand{\arraystretch}{1.25}
\begin{deluxetable}{ccccl}[t]
	\tablewidth{0.95\textwidth}
	\tablecolumns{5} 
	\tablecaption{Variability Statistics\label{table:varstat}}
	\tablehead
	{
		\colhead{Band}&\colhead{Central Wavelength}&\colhead{$\sigma_{m}$}&\colhead{$R_\mathrm{max}$}&\colhead{$\fvar$}
		\\
		\colhead{}&\colhead{$\mathrm{(\mu m)}$}&\colhead{(mag)}&\colhead{}&\colhead{(\%)}
	}
	\startdata 
\multicolumn{5}{c}{2017}\\
UVM2 	& 0.225& 0.13& 1.87& $8.1\pm1.2$ 		\\
V		& 0.551& 0.10& 1.70& $8.2\pm0.1$   	\\
J 	 	& 1.22 & 0.03& 1.15& $2.7\pm0.2$   	\\
H 	 	& 1.63 & 0.02& 1.08& $1.6\pm0.3$    	\\
K 	 	& 2.19 & 0.12& 1.37& $1.1\pm0.1$    	\\
\ha\tablenotemark{$a$} & 0.66 & 0.01 & 1.07 & $0.6\pm 0.1$
\\
\hline
\multicolumn{5}{c}{2018}\\
V 						& 0.551 & 0.05 & 1.23& $4.7\pm0.1$ \\
\ha\tablenotemark{$b$}	& 0.66  & 0.02 & 1.08 & $1.2\pm 0.1$\\
\hline
\multicolumn{5}{c}{2017-04-28}\\
J 		& 1.22 & 0.03 & 1.12& $2.4\pm0.2$	\\
H 		& 1.63 & 0.02 & 1.08& $1.6\pm0.3$	\\
K 		& 2.19 & 0.02 & 1.07& $1.2\pm0.6$   \\
\hline
\multicolumn{5}{c}{2017-05-02}\\
V				 		& 0.551& 0.03 & 1.11 & $2.5\pm0.1$   \\
\ha\tablenotemark{$b$} 	& 0.657& 0.01 & 1.04 & $0.1\pm0.4$
\\
\hline
\multicolumn{5}{c}{2018-04-08}\\
V 						& 0.551& 0.02 & 1.10 & $1.7\pm0.1$    \\
\ha\tablenotemark{$b$}	& 0.657& 0.01 & 1.03 & $0.6\pm0.1$
	\enddata
	\tablenotetext{a}{Light curves from MDM 1.3m \& MDM 2.4m}
	\tablenotetext{b}{Light curves from MDM 2.4m}
\end{deluxetable}

\subsection{Continuum Correction for \ha\ Photometry}\label{ss:contcorr}

Ideally, spectroscopic monitoring can provide better data to measure the \Ha\ emission-line flux by separating the emission line from the continuum based on the spectral decomposition, leading to less uncertainty in the cross-correlation analysis between AGN continuum and \Ha\ emission line. The main uncertainty of photometric reverberation mapping comes from the fact that the contribution from the AGN continuum to the total flux obtained with a broad filter has to be properly determined \citep{Desroches+06}. Compared to a broad-band filter, the narrow \Ha-band filter contains less AGN continuum and can be effectively used for the \Ha\ emission-line flux monitoring. If the continuum contribution in the narrow-band \Ha\ filter can be properly removed, narrow-band photometry can lead to successful measurements of the \Ha\ emission-line flux. In this section, we investigate the effect of the variability of the continuum in the \Ha\ band. 

\subsubsection{Test of Continuum Variability for \ha\ Photometry}\label{ss:contvar}

The total flux measured with a narrow-band \ha\ filter is composed of the flux from the broad \ha\ line, narrow emission lines, and the continuum emission from the AGN and its host galaxy. Thus, we model the narrow-band \ha\ flux $F_\mathrm{nH\alpha} (t)$ as
\begin{align}
F_\mathrm{nH\alpha} (t) = f_\mathrm{BH\alpha} (t) + f_\mathrm{cont} (t) + f_\mathrm{NL},
\end{align}
where $f_\mathrm{BH\alpha} (t)$ is the variable flux of the broad \ha\ emission, $f_\mathrm{cont}(t)$ is the variable flux of the continuum from both AGN and nonvarying stars, and $f_\mathrm{NL}$ is the nonvarying flux from narrow emission lines. While the variability of $f_\mathrm{BH\alpha} (t)$ is delayed with respect to the V-band continuum, the variability of $f_\mathrm{cont}(t)$ is similar to that of V. If we ignore the difference of the wavelength between V and \Ha, the flux variability of $f_\mathrm{cont}(t)$  is to be coherent with that of the V band. This assumption is reasonable if there is no color variability in this relatively short wavelength range covered by the V and \Ha\ bands. Furthermore, if there is no significant contribution from nonvarying stars to the \Ha\ filter, then the variability amplitude of the continuum will be similar between the V-band and \Ha-band spectral ranges. 

A key for the proper continuum correction is to have a high-quality spectrum with which the AGN continuum fraction can be reliably determined. We used the mean spectrum obtained with the Gemini GMOS during our 3.5\,hr observing run for measuring the flux contribution of each component based on the spectral decomposition. We modeled the GMOS spectrum with multiple components: the narrow \ha\ emission line and \NII\ $\lambda\lambda 6548$, 6583 doublet, the broad \ha\ emission line, and the continuum. We used double Gaussian models for the narrow emission line to account for the core and wing components, a single Gaussian component for the broad \Ha\ emission line, and a first-order polynomial for the continuum (see Figure~\ref*{fig:contfrac}). After convolving each component with the response function of the  KP1468 \Ha\ filter, we determined that the continuum is $18.3 \pm 0.3$\% of the total flux in the narrow-band \Ha\ filter, while the narrow line and broad line contribute 49\% and 32\% of the flux, respectively.

\begin{figure}
	\center
	\includegraphics[width=0.49\textwidth]{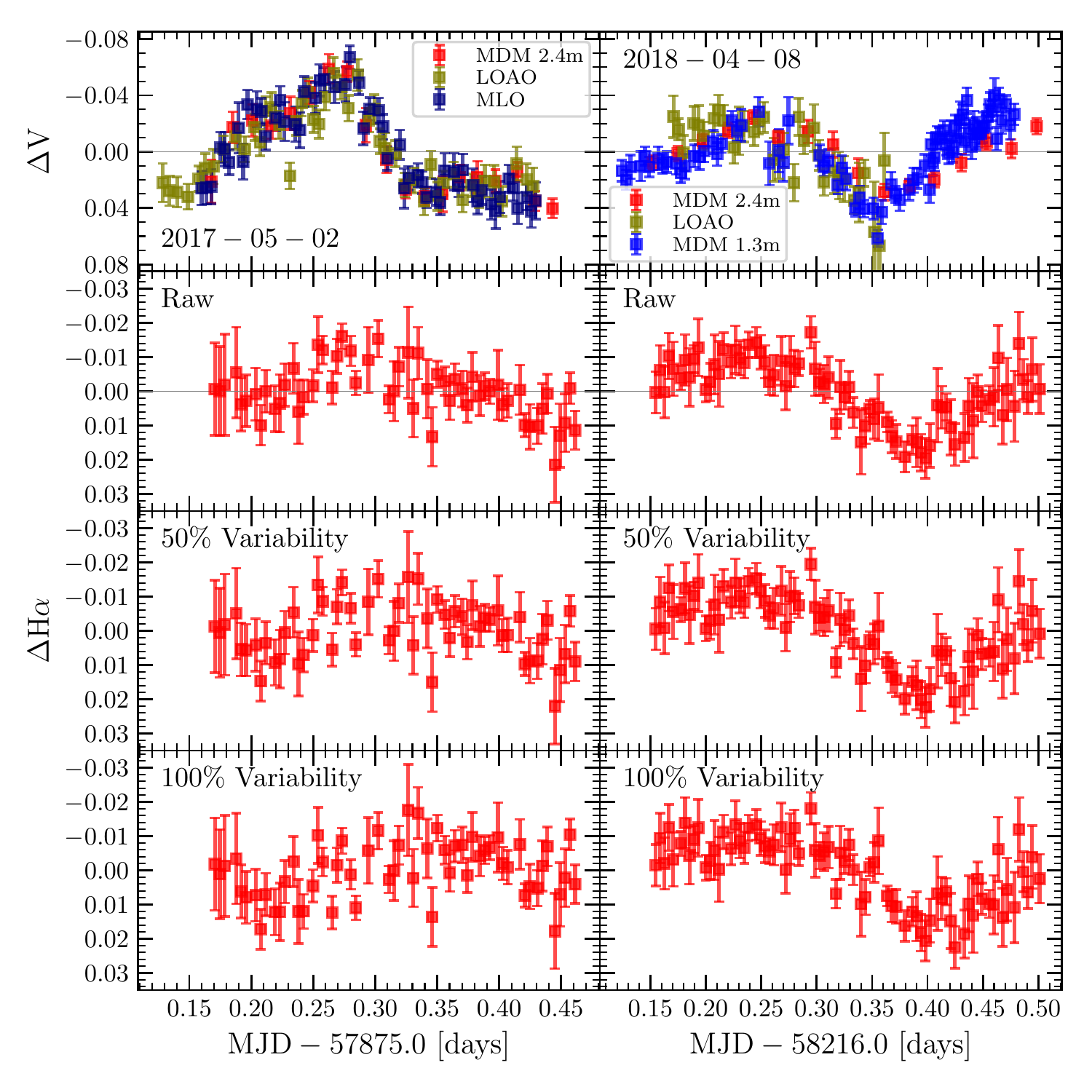}
	\caption{
		Light curves in V (top) and the narrow \ha\ filter (bottom 3 panels) obtained on 2017-05-02 (left) and 2018-04-08 (right), using the MDM telescopes, MLO, and LOAO. \ha\ light curves with continuum subtracted are also presented, assuming a continuum fraction of 18.3\% of the total \ha\ flux on average and variability of 50\% and 100\% of that of the V-band light curves. \label{fig:lc}} 
\end{figure}

On the other hand, the variability amplitude of the continuum in the narrow-band \ha\ filter can differ from that of the V band if the variability amplitude depends on the continuum wavelength, while it is reasonable to assume the same amplitude, considering the small difference of the wavelengths between two filters. We model the continuum in the \Ha\ filter by quantifying the dimensionless variability amplitude $K$ as 
\begin{align}
f_\mathrm{cont} (t) = \left[(1-K) + K \frac{f_\mathrm{V} (t)}{\left\langle f_\mathrm{V} \right\rangle} \right] \left\langle f_\mathrm{cont}\right\rangle, \label{eq:havari}
\end{align}
where $\left\langle f_\mathrm{cont} \right\rangle$ is the mean continuum flux in the \Ha\ filter, $\left\langle f_\mathrm{V} \right\rangle$ is the mean flux of the V-band filter, and $f_\mathrm{V}$ (t) is the V-band flux at each epoch. For example, if $K=1$, then the variability amplitude of the continuum is the same between the V band and \Ha, while the continuum in the \Ha\ band has no variability for $K=0$. Thus, by parameterizing the variability amplitude by $K$, we can test the effect of the continuum contribution to the \Ha\ light curve. Different assumptions on variability amplitude would affect the recovered \ha\ light curve as shown in Figure~\ref{fig:lc}, where we present the change of the narrow-band \Ha\ filter light curve using $K=0,\, 0.5,$ and 1.

\subsection{The Effect of Continuum Contribution on the Continuum-\ha\ Time Lag}\label{ss:lag}

To quantify the effect of the continuum contribution in the \Ha-band on the time-lag analysis, we used five \Ha\ light curves, which were corrected for the continuum contribution with an assumed variability amplitude ($K$ in Eq.~\eqref{eq:havari}). 

For measuring a time lag from a pair of light curves (V and \Ha), we computed the interpolated cross-correlation function (ICCF; \citealt{White&Peterson94}). We adopted the flux randomization/random subset selection method (FR/RSS; \citealt{Peterson+98}; see also \citealt{Peterson+04}) to estimate its uncertainty. The ICCF $r(\tau)$ was computed over $-4\,\hr<\tau<4\,\hr$ with $0.01\,\hr$ interval. We obtained two ICCFs by interpolating either the V-band light curve or the \Ha\ light curve and then adopted the average of the two ICCFs. We simulated 2000 realizations with the FR/RSS. For each realization, we resampled each light curve by allowing any epoch to be drawn multiple times. Then the flux at each epoch was randomized with a log-normal distribution corresponding to the measured flux, flux uncertainty, and the number of times that epoch was drawn in the resampling step. The centroid of the ICCF, defined as the ICCF-weighted mean of $\tau$ where $r(\tau)>0.8\, r_\mathrm{max}$, was calculated for each realization. Finally, the median and the lower/upper bounds of the 68\% central confidence interval of the centroid distribution were taken as the time lag and its lower/upper uncertainty. In addition, we used the $z$-transformed discrete correlate function (\zdcf;  \citealt{Alexander97}) and the \jav\ method \citep{Zu+11} to measure the time lag between V-band and \ha-band light curves, in order to compare with the ICCF results. Our measurements of continuum-\ha\ time lag are summarized on Table \ref{table:timelag}.

\renewcommand{\arraystretch}{1.}
\begin{deluxetable}{ccccc}[t]
	\tablewidth{0.9\textwidth}
	\tablecolumns{5} 
	\tablecaption{Time Lags for Broad \ha\ Line from Photometric Light Curves \label{table:timelag}}
	\tablehead
	{
	\colhead{Date}  &\colhead{Continuum correction}&\colhead{$\tau_{\rm ICCF}$}&\colhead{$\tau_{\rm zDCF}$}&\colhead{$\tau_{\rm JAV}$}\\        
	\colhead{(UT)}  &                                                    &\colhead{(min)}      &\colhead{(min)}&\colhead{(min)}\\ 
	\colhead{(1)}    &\colhead{(2)}       &\colhead{(3)}       &\colhead{(4)}&\colhead{(5)}   
	}
	\startdata
	2017-05-02 & no correction  & \valerrud{55}{27}{31} & \valerrud{ 67}{22}{32} & \valerrud{59}{14}{14} \\
	&  25\% of V-band variability & \valerrud{72}{25}{33} & \valerrud{ 67}{36}{21} & \valerrud{74}{18}{14} \\
	&  50\% of V-band variability  & \valerrud{88}{27}{44} & \valerrud{119}{27}{37} & \valerrud{98}{17}{22} \\
	&  75\% of V-band variability   & \valerrud{104}{31}{55} & \valerrud{119}{30}{20} & \valerrud{120}{14}{22} \\
	& 100\% of V-band variability  & \valerrud{122}{33}{67} & \valerrud{147}{20}{32} & \valerrud{135}{7}{52} \\
\hline
	2018-04-08 & no correction & \valerrud{49}{15}{14} & \valerrud{ 33}{24}{27} & \valerrud{68}{11}{22}\\
	&  25\% of V-band variability & \valerrud{56}{13}{13} & \valerrud{ 67}{ 4}{32} & \valerrud{76}{36}{11} \\
	&  50\% of V-band variability  & \valerrud{64}{14}{14} & \valerrud{ 67}{23}{28} & \valerrud{84}{28}{13} \\
	&  75\% of V-band variability & \valerrud{73}{14}{14} & \valerrud{ 67}{30}{23} & \valerrud{94}{29}{13} \\
	&  100\% of V-band variability  & \valerrud{83}{13}{14} & \valerrud{ 99}{ 9}{35} & \valerrud{100}{18}{11} \\
\hline
\hline	
	2017-04-29 & no correction & \valerrud{79}{30}{25} & \valerrud{ 70}{11}{16}& - \\
	2017-04-30 & no correction & \valerrud{84}{86}{66} & \valerrud{131}{29}{80} &- \\
	2017-05-01 & no correction & \valerrud{2}{19}{14} & \valerrud{  8}{21}{16} & - \\
	\enddata
	\tablecomments{Rest-frame time-lag measurements after subtracting the continuum contribution from the \ha\ narrow-band flux. The continuum flux is on average 18.3\%, but varies as in the V-band variability. The time lag represents the median of the distribution for ICCF and \jav, and the maximum likelihood lags for \zdcf. Central 68\% intervals are taken as their uncertainties.}
\end{deluxetable}

\begin{figure*}
	\center
	\includegraphics[width=0.49\textwidth,]{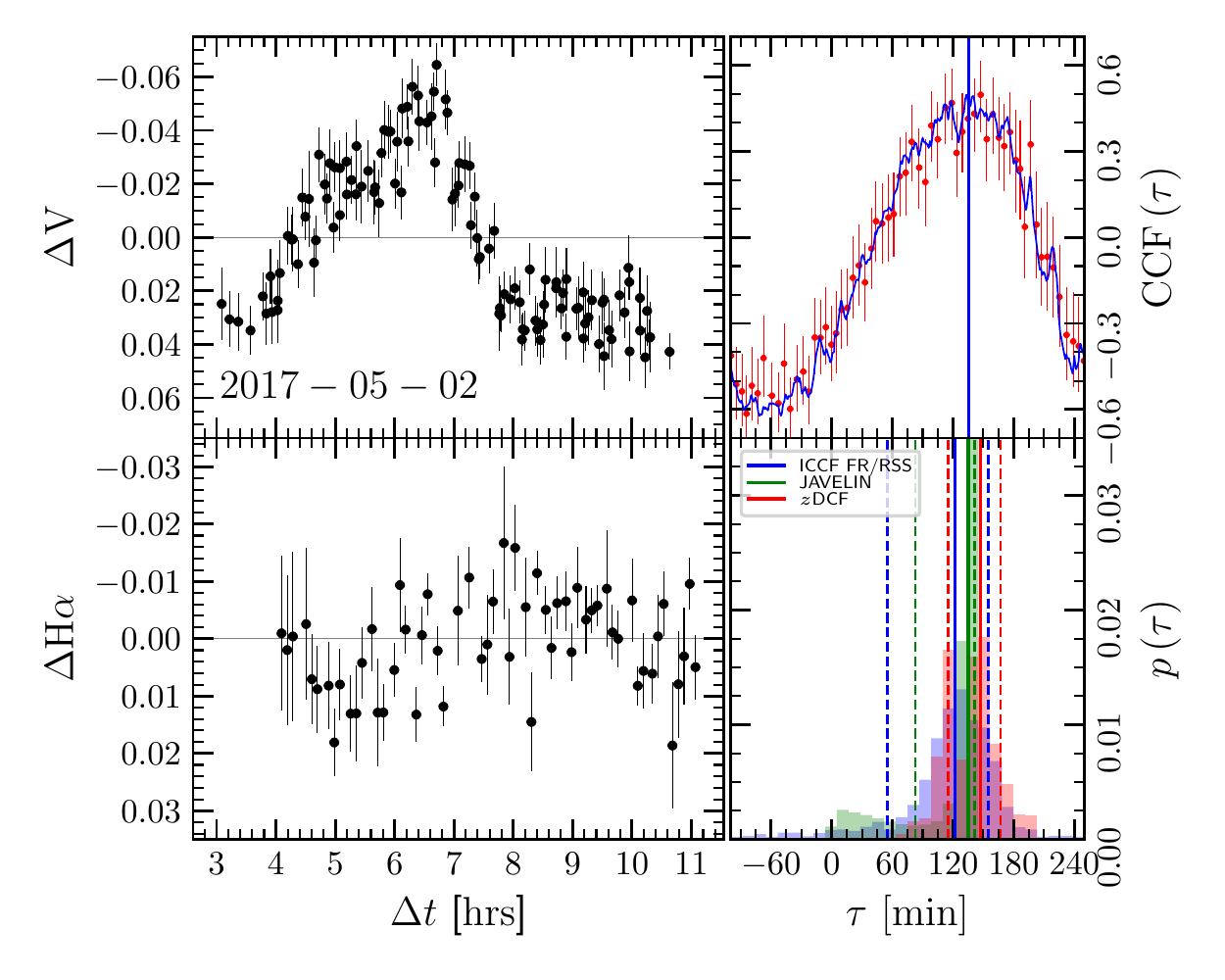}
	\includegraphics[width=0.49\textwidth]{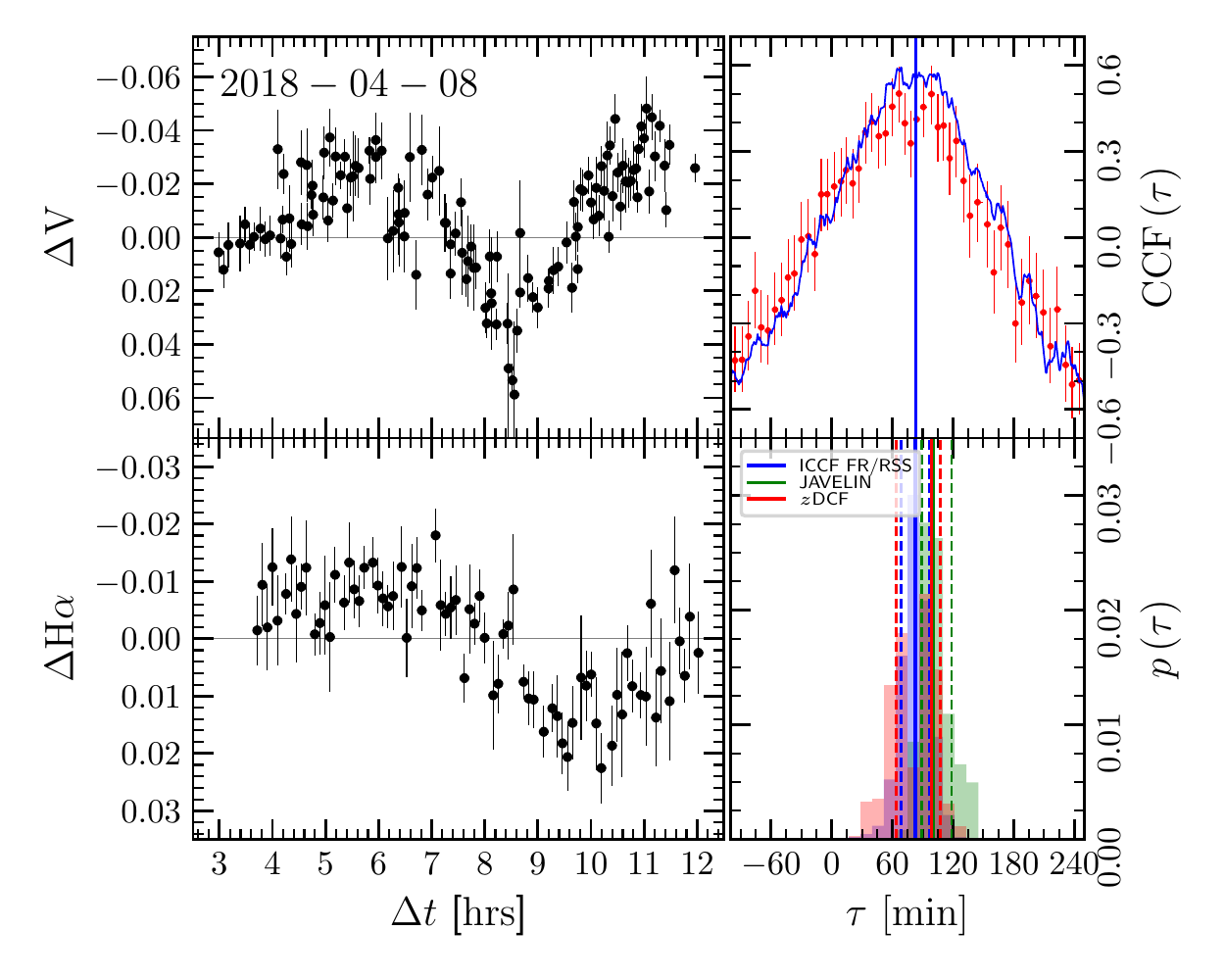}
	\caption{
		Light curves and corresponding ICCF centroid and \jav\ results, using the data from 2017-05-02 (\emph{left set}) and 2018-04-08 (\emph{right set}). Each set of figures consists of 4 panels as follows. 
		\emph{Left}: V (\emph{top}) and \ha\ (\emph{bottom}) light curves, where \ha\ light curve is after correction for continuum contamination described in \ref{ss:contcorr}. 
		\emph{Top Right}: ICCF (blue) and \zdcf\ (red) of the data, where the ICCF centroid is represented as a vertical line.
		\emph{Bottom Right}: Probability distributions of ICCF centroids (blue), \jav\ models (green), and \zdcf\ (red) for the data.
		Solid vertical lines mark the median (for ICCF and \jav ) or maximum likelihood lag (for \zdcf) with dashed lines marking their central 68\% intervals.
		\label{fig:ccf2017}
	}
\end{figure*}

First, we present the ICCF, \zdcf, and \jav\ measurements using the best light curves from 2018-04-08 in Figure \ref*{fig:ccf2017}. We also used the light curves from 2017-05-02 for a consistency check. Without correcting for the continuum contribution to the \ha-band, we obtained the ICCF time lag \valerrud{55}{27}{31} min from the 2017-05-02 data and \valerrud{49}{15}{14} min for the 2018-04-08 data. These results are consistent with those of \zdcf{} and \jav{} measurements, where \valerrud{67}{22}{32} min (\zdcf{}) and \valerrud{59}{14}{14} min (\jav{}) were measured for the 2017-05-02 data and \valerrud{33}{24}{27} min (\zdcf{}) and \valerrud{68}{11}{22} min (\jav{}) were measured for the 2018-04-08 data.

Although the quality of the light curves is much lower, we also tried to measure the time lag using the light curves from other dates, including 2017 April 29, April 30, and May 1. As summarized in Table~\ref{table:timelag}, the obtained time lag from these dates suffers large uncertainty owing to the poor data quality and the limited time baseline, and the lack of strong variable features. Nevertheless, we find that the lag measurements are broadly consistent with that of the best light curves from 2018-04-08. 

Second, we used the light curves from the best two dates to test the effect of the continuum variability in the \Ha-band on the time lag measurement. Assuming the variability amplitude of the continuum in the \Ha-band is 25\%, 50\%, 75\%, and 100\% of that of V, we subtracted the continuum contribution from the total flux observed with the \Ha{} filter, which was on average 18.3\%, but slightly varied as the V-band light curve. The continuum flux observed through the \Ha{} filter generally decreases the time lag between the V-band continuum and \Ha\ line since the continuum in the V-band as well as in the \Ha-band has the same variability pattern. Thus, if we assume a smaller variability amplitude of the continuum, the continuum variability signal is less subtracted from the \Ha-band light curve, weakening the variability pattern of the \Ha\ emission-line flux. For example, we can obtain a lower limit of the lag if we do not correct for the continuum variability (i.e., assuming 0\% variability amplitude) in the \Ha-band light curve. 

As shown in Figure~\ref{fig:contvar}, the time lag increases by almost a factor of two with the maximum correction ($K=1$). While the three analysis methods (ICCF, zDCF, and JAVELIN) provided somewhat different lag measurements, they are mostly consistent within the uncertainties. Using the light curves from 2017-05-02, we obtained consistent results, with an increasing time lag with the higher variability amplitude. Table~\ref{table:timelag} summarizes the time-lag measurements, depending on the used light curves and the analysis method. 

\begin{figure}
	\center
	\includegraphics[width=0.48\textwidth, height=7cm]{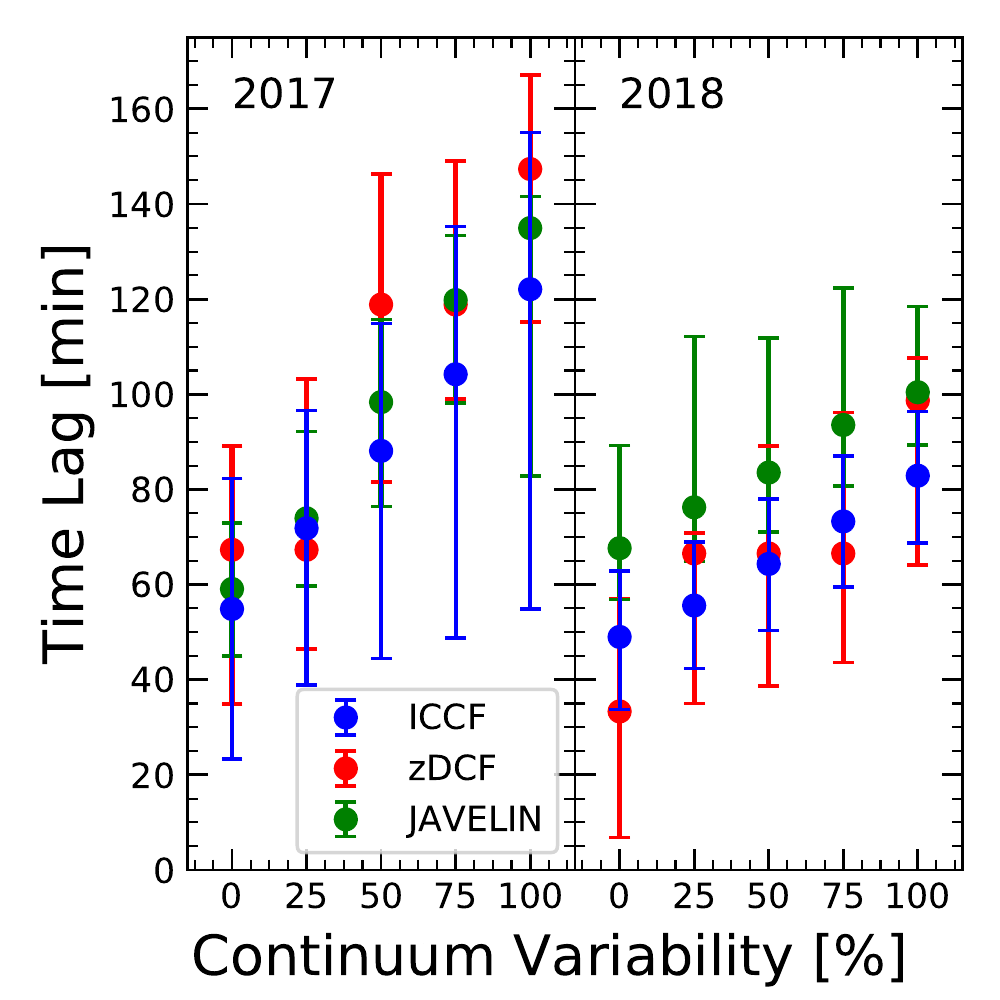}
	\caption{{Effects of continuum variability on time-lag determination.} Time lags determined from ICCF (blue), \zdcf\ (red), and \jav\ (green) when the continuum variabilities in the \ha\ narrow-band light curves are assumed to be some fraction of that of the V band for 2017 (\emph{Left}) and 2018 (\emph{Right}). Continuum fraction is assumed to be 18.3\%. \label{fig:contvar}} 
\end{figure}

Assuming the variability amplitude is the same between V and \Ha, we compensated for 18.3\% continuum contribution in the \ha\ light curve, and we found the ICCF \ha\ time lag to be $\tau_\mathrm{ICCF}=$\valerrud{122}{33}{67} min on 2017-05-02 and $\tau_\mathrm{ICCF}=83\pm 14$ min on 2018-04-08, which are consistent within 1$\sigma$ uncertainties. We also conducted ICCF measurements with \ha\ light curves compensating for 18.0\% and 18.6\% continuum, which are the 1$\sigma$ bounds for the continuum fraction in the narrow-band \ha\ filter. We found that the difference in time lag is not larger than 1 min; thus, the uncertainty arising from the error in continuum contamination measurement can be ignored. 

On the other hand, we did find a difference in the time-lag measurement if the variability amplitude of the continuum in the \ha\ filter is different from that of the V-band filter, as shown in Figure~\ref{fig:contvar}. We also checked the consistency between ICCF and \zdcf\ or \jav\ time lags. We found that \zdcf\ time lags are consistent with ICCF lags within 1$\sigma$. \jav\ time lags seem to be systematically larger than those of ICCF and \zdcf, although the difference is within 1$\sigma$. 

\begin{figure}
	\center
	\includegraphics[width=0.49\textwidth]{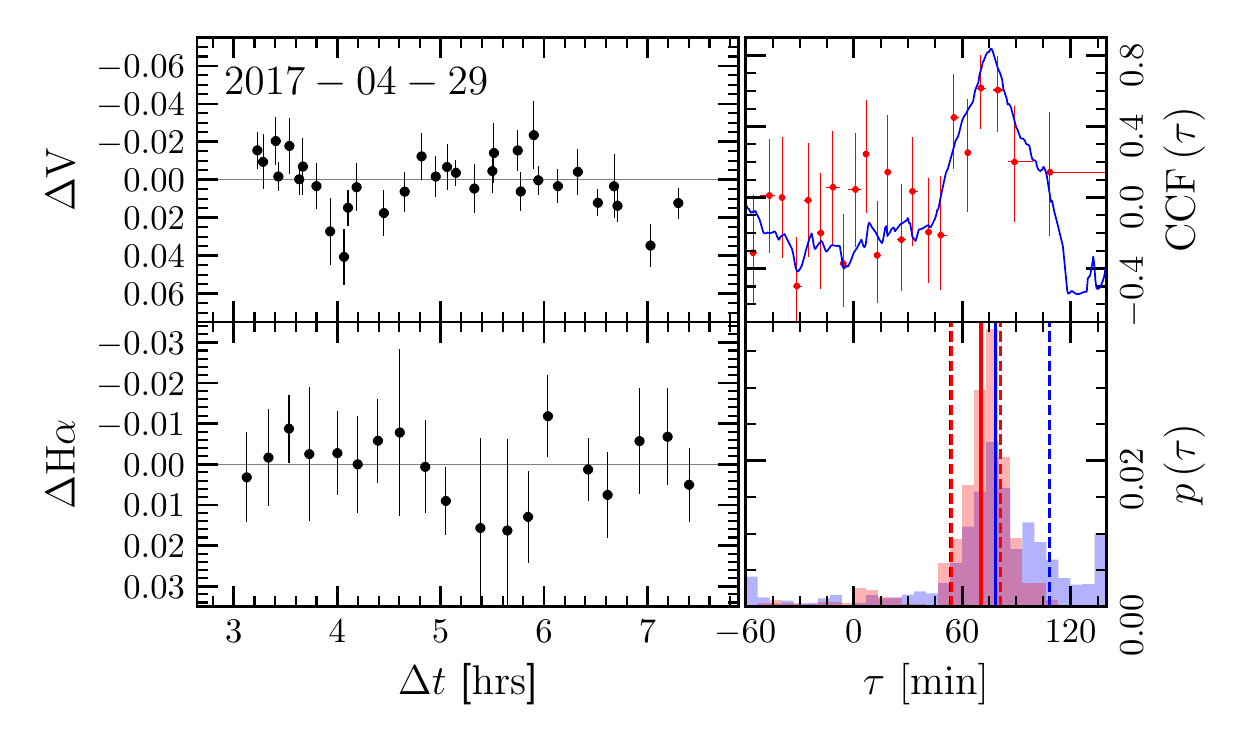}
	\includegraphics[width=0.49\textwidth]{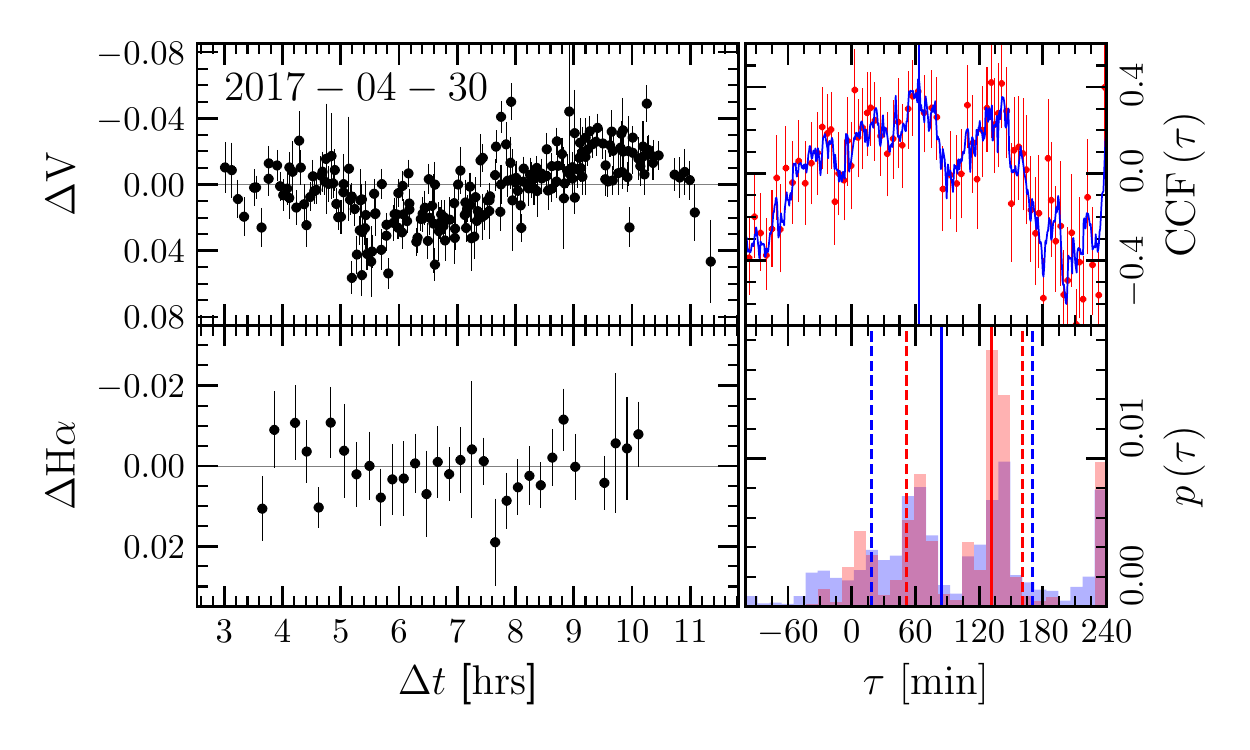}
	\includegraphics[width=0.49\textwidth]{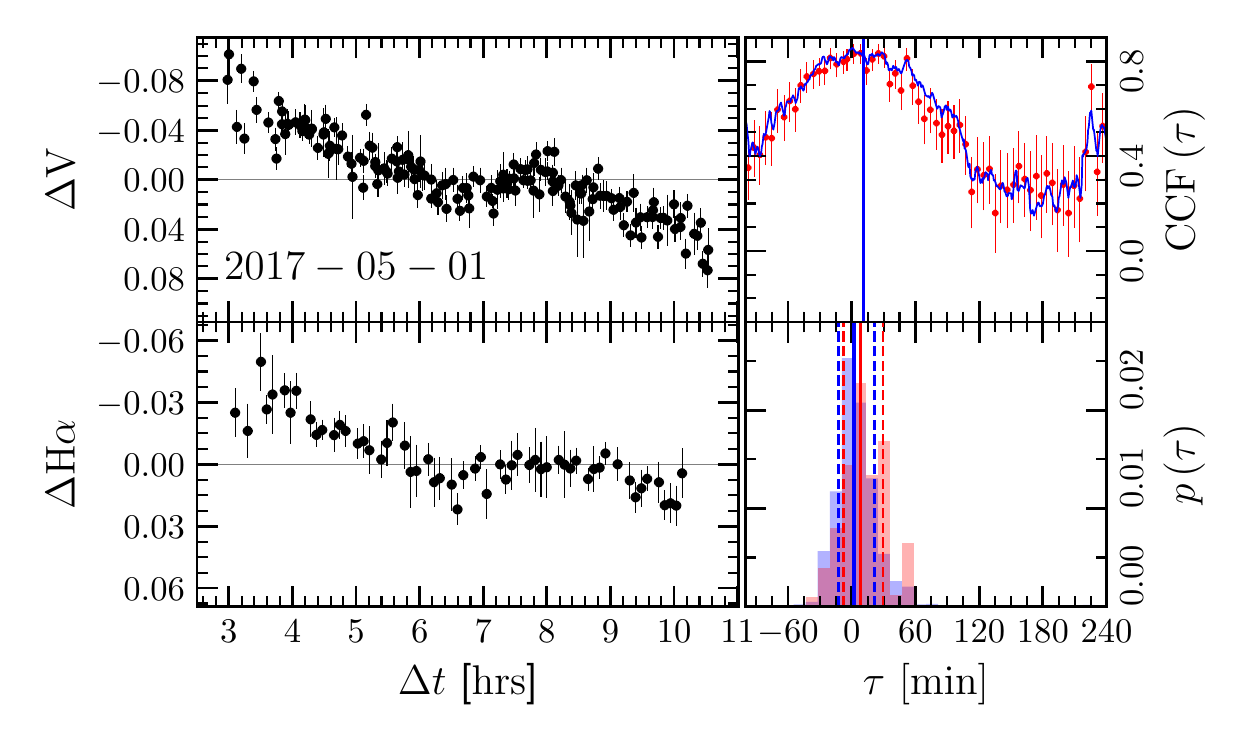}
	\caption{ICCF between V-band and raw narrow-band \ha\ light curves. From top to bottom, each set of figures shows the ICCF analysis result for 2017-04-29, 2017-04-30, and 2017-05-01 without correcting for continuum variability in \ha\ light curves as follows. \emph{Left}: V (top) and \ha\ (bottom) light curves, where the \ha\ light curve is without correction for continuum contamination. 
		\emph{Top Right}: ICCF of the data, where the centroid is represented as a vertical line.
		\emph{Bottom Right}: Probability distributions of ICCF centroids (blue) and \zdcf\ (red) for the data. 
		Solid vertical lines mark the median (for ICCF) or maximum likelihood lag (for \zdcf), with dashed lines marking their central 68\% intervals.
		\label{fig:ccf_raw}
	}
\end{figure}

\begin{figure}
	\center
	\includegraphics[width=0.49\textwidth]{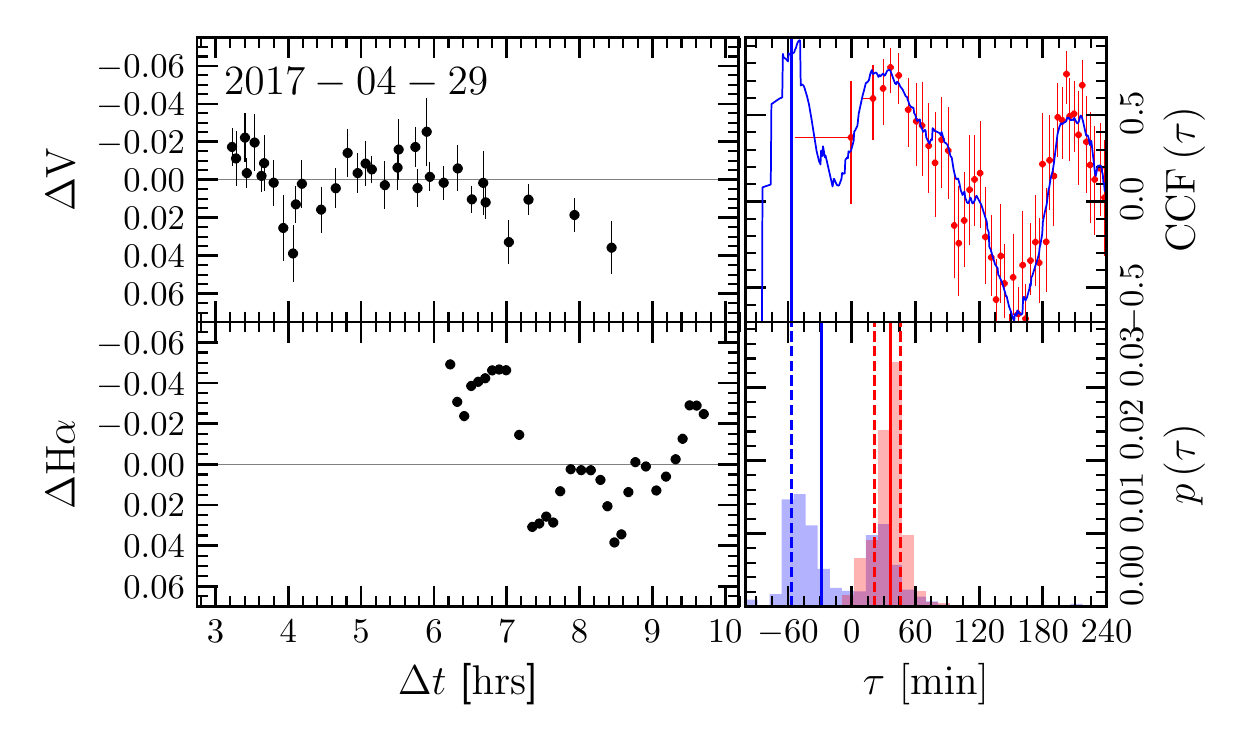}
	\caption{ICCF between the V band and spectroscopic broad \ha\ emission-line flux, normalized to the \SII\ narrow lines. \emph{Left}: V (top) and broad \ha\ (bottom) light curves. Note that \ha\ line flux is converted into relative magnitudes.
		\emph{Top Right}: ICCF of the data, where the centroid is represented as a vertical line.
		\emph{Bottom Right}: Probability distributions of ICCF centroids (blue) and \zdcf\ (red) for the data.
		Solid vertical lines mark the median (for ICCF) or maximum likelihood lag (for \zdcf), with dashed lines marking their central 68\% intervals. Note that this result is unreliable owing to the flux uncertainties caused by bad weather and the limited time baseline (see \ref{ss:lag}).
		\label{fig:ccf_gemini}
	}
\end{figure}

As a consistency check for the \Ha\ time lag, we also used the light curves from 3 nights: 2017-04-29, 2017-04-30, and 2017-05-01. Since these light curves showed relatively large flux uncertainties due to bad weather, we did not correct for the continuum contribution and measured the time lag of the \Ha-band light curve with respect to the V-band light curve as shown in Figure \ref*{fig:ccf_raw}. For these measurements, we only used the \Ha-band data from the MDM 1.3m by excluding the low-quality data with large uncertainties from the BOAO 1.8m.  The measured time lags from 2017-04-29 and 2017-04-30 have large uncertainties as expected and are roughly consistent with the best lag measurement from 2018-04-08 within the error. In the case of 2017-05-01, we obtained a time lag consistent with zero, presumably due to the lack of a strong pattern in the light curves. 

During the campaign on 2017-04-29, we obtained spectroscopic monitoring data with the Gemini GMOS for $\sim$3.5\,hr and constructed a light curve of the \Ha\ emission line. By cross-correlating with the V-band light curve, we measured the time lag as shown in Figure \ref*{fig:ccf_gemini}. Note that the flux calibration has large uncertainties since the sky conditions were quickly changing during the campaign, which had to be ended after $\sim$3.5\,hr. We calibrated the flux of the broad \ha\ emission line, by assuming that the \SII\ emission-line flux is constant. Then we converted the line flux to magnitude units for consistency with the \Ha-band light curves. We could not obtain meaningful results since the overlap between the V band and the \Ha\ light curves was limited, as was the sampling.

\subsection{UV-to-IR Continuum Time Lag}\label{ss:cclg}

We investigated the time lag between continuum bands using the UV, optical, and near-IR light curves. While all continuum light curves showed consistent variability patterns, we were not able to detect any reliable lag between two continuum bands, as summarized in Table \ref{table:othertimelag}. 

First, we performed a cross-correlation analysis using the UV and V-band light curves. However, these light curves have several limitations. The UV light curve has gaps of approximately an hour between epochs owing to the invisibility of the target in each orbit of \swift. Thus, a relatively short lag of $\sim 1$\,hr is challenging to measure. 
In addition, the flux uncertainties of the UVM2 band are relatively high, $\Delta m \approx 0.05$--0.1, comparable to the fractional variability of the UVM2 light curve. 

Second, we investigated the lag of the near-IR continuum. Unfortunately, there was no time baseline when the optical and near-IR monitoring observations were performed simultaneously. Thus, we only compared among J, H, and K-band light curves. We obtained no meaningful time-lag measurements among the J, H, and K light curves as their temporal baselines were relatively short, and the sampling and time resolution were limited (see Figure~\ref{fig:lc2017}). 

\begin{deluxetable*}{ccccccc}[t]
	\tablewidth{0.9\textwidth}
	\tablecolumns{6} 
	\tablecaption{Time Lags Between Continuum Light Curves \label{table:othertimelag}}
	\tablehead
	{
		\colhead{Date}&\colhead{Telescopes}&\colhead{Light Curve 1}&\colhead{Light Curve 2} &\colhead{$\tau$}&\colhead{Method}\\        
		\colhead{(UT)}&  &         &         &\colhead{(min)}      &\\ 
		\colhead{(1)} &\colhead{(2)}       &\colhead{(3)}       &\colhead{(4)}&\colhead{(5)}&\colhead{(6)}
	}
	\startdata
	2017-04-30& UVOT, BOAO, DOAO, LOAO, & UVM2 & V & \valerrud{66}{277}{146} & ICCF\\
	&MDM 1.3m, MLO, Nickel, WMO &&&\valerrud{-136}{157}{11}& \zdcf \\
	\\
	2017-05-01& UVOT, BOAO, DOAO, Hiroshima, & UVM2 & V & \valerrud{135}{134}{126} & ICCF \\
	&LOAO, MDM 1.3m, MLO, Nickel &&&\valerrud{228}{15}{259}& \zdcf \\
		&&&&&\\
	2017-04-26& CMO 2.5m & J & K & \valerrud{0}{95}{119} & ICCF \\
		&&&&\valerrud{-7}{24}{7}& \zdcf \\
	\\
	2017-04-28& CMO 2.5m & J & H & \valerrud{-7}{91}{92} & ICCF\\
		&&&&\valerrud{-33}{47}{15}& \zdcf\\
	\\
	2017-04-28& CMO 2.5m & J & K & \valerrud{2}{129}{145} & ICCF\\
		&&&&\valerrud{34}{16}{13}& \zdcf\\
	\\
	2017-04-28& CMO 2.5m & H & K & \valerrud{24}{122}{158} &ICCF\\
		&&&&\valerrud{47}{177}{20}&\zdcf\\
	\enddata
	\tablecomments{Rest-frame time lag values are chosen from the medians of the distributions for ICCF, and from the maximum likelihood lags for \zdcf. Central 68\% intervals are taken as their uncertainties.}
\end{deluxetable*}

\subsection{AGN Luminosity and the BLR Radius-Luminosity Relation}\label{ss:lum}

In this section, we investigate the size-luminosity relation by measuring the monochromatic luminosity at 5100\,\AA. First, we calculate the mean V-band magnitude of the AGN from the light curve on 2018-04-08 and obtain $\mathrm{V_{AGN}}=16.1$. Second, we rescale the mean spectrum constructed from the 3.5\,hr GMOS observations on 2017-04-29 by multiplying a scale factor of 0.507, in order to match the synthetic V-band magnitude measured from the mean spectrum with the photometry result $\mathrm{V_{AGN}}=16.1$\,mag. Here, we assume that the spectral shape changed insignificantly during 2017 and 2018. Then, we measure the monochromatic luminosity at 5100\,\AA\ from the rescaled mean spectrum and obtain $\lumcont = 1.02\times 10^{40}\, \mathrm{erg\,s^{-1}}$, after Galactic extinction correction, which was adopted as 0.05\,mag from \cite{Carson+15}.

Note that the determined $\lumcont$ is an upper limit of the AGN luminosity since there is a contribution from the host-galaxy stellar component. To investigate the effect of host-galaxy contribution, we model the radial surface brightness profile of the central source using two components: a point source and an exponential disk. For this analysis, we construct an average image using the high-quality V-band image data (i.e.,$16 \times 180$\,s exposure), which were obtained with the MDM 2.4m telescope on 2018-04-08. Note that we use the images from the same date, from which we measured the time lag of the \Ha\ emission line, in order to secure consistent measurements of the luminosity and the lag.  Using the same aperture size as for the aperture photometry, we calculate the flux from a point source to be 84\% of the total flux in the aperture. If we remove the 16\% contribution from the host galaxy disk, then the AGN luminosity becomes $\lumcont = 8.52\times 10^{39}\, \mathrm{erg\,s^{-1}}$.

However, a more serious issue in measuring AGN monochromatic luminosity is the presence of a nuclear star cluster (NSC), which is a point-like source with an effective radius $< 0.3''$ \citep{Carson+15}. Since we cannot separate the NSC and the AGN in our images with a large seeing disk, we instead adopt the estimated luminosity of the NSC based on the modeling of the SED from \cite{Carson+15}, which is \valerrud{-9.78}{0.03}{0.04} mag in the {F438W} band and \valerrud{-10.48}{0.06}{0.09} in the {F547M} band after Galactic extinction correction, and determine the luminosity at 5100\,\AA. After correcting the luminosity for the distance to NGC 4395 that we adopted in this work, we obtain $\lumct{\mathrm{NSC}}=3.59\times 10^{39}\, \mathrm{erg\,s^{-1}}$. By subtracting the flux from the NSC, we determine the AGN luminosity to be $\lumct{\mathrm{AGN}}=5.75\times 10^{39}\, \mathrm{erg\,s^{-1}}$. 

To estimate the uncertainty of the AGN luminosity we include several sources of error: (1) the flux-measurement uncertainty at 5100\,\AA\ is 1.54\%, (2) the systematic uncertainty due to the conversion of the SDSS magnitudes of comparison stars to the V band magnitudes is 0.01\,mag \citep{Jester+05}, (3) the standard deviation of the mean magnitude from the V-band light curve is 0.015\,mag, and (4) the flux-measurement uncertainty of the NSC is 0.075\,mag \citep{Carson+15}. Combining these errors, we determine $\log \lumct{\mathrm{5100_{AGN}}} = 39.76\pm0.03$, or $\lumct{\mathrm{5100_{AGN}}} = \left(5.75\pm0.40\right)\times 10^{39}\,\mathrm{erg\,s^{-1}}$. 

By combining the monochromatic luminosity at 5100\,\AA\ and the best measurement of the \Ha\ lag, $\tau=83\pm 14$ min, we compare NGC 4395 with other reverberation-mapped AGNs in the size-luminosity relation (Figure~\ref{fig:sl}). NGC 4395 is offset by 0.48\,dex from the size-luminosity relation defined by more luminous AGNs \citep[][case for \emph{Clean+ExtCorr} in Table 14]{Bentz+13}. If we consider the intrinsic scatter (i.e., $\lesssim 0.19$\,dex) of the relation reported by \cite{Bentz+13}, the offset is significant ($\geq 2.5\sigma$). On the other hand, the systematic uncertainty of the AGN luminosity of NGC 4395 can be very large due to the difficulty of separating the AGN from the NCS.
Note that recent reverberation studies showed that more luminous AGNs are scattered below the best-fit relation given by \citet{Bentz+13}; see \S\ref{ss:sl} for more details. 

\begin{figure}
	\center
	\includegraphics[width=0.45\textwidth]{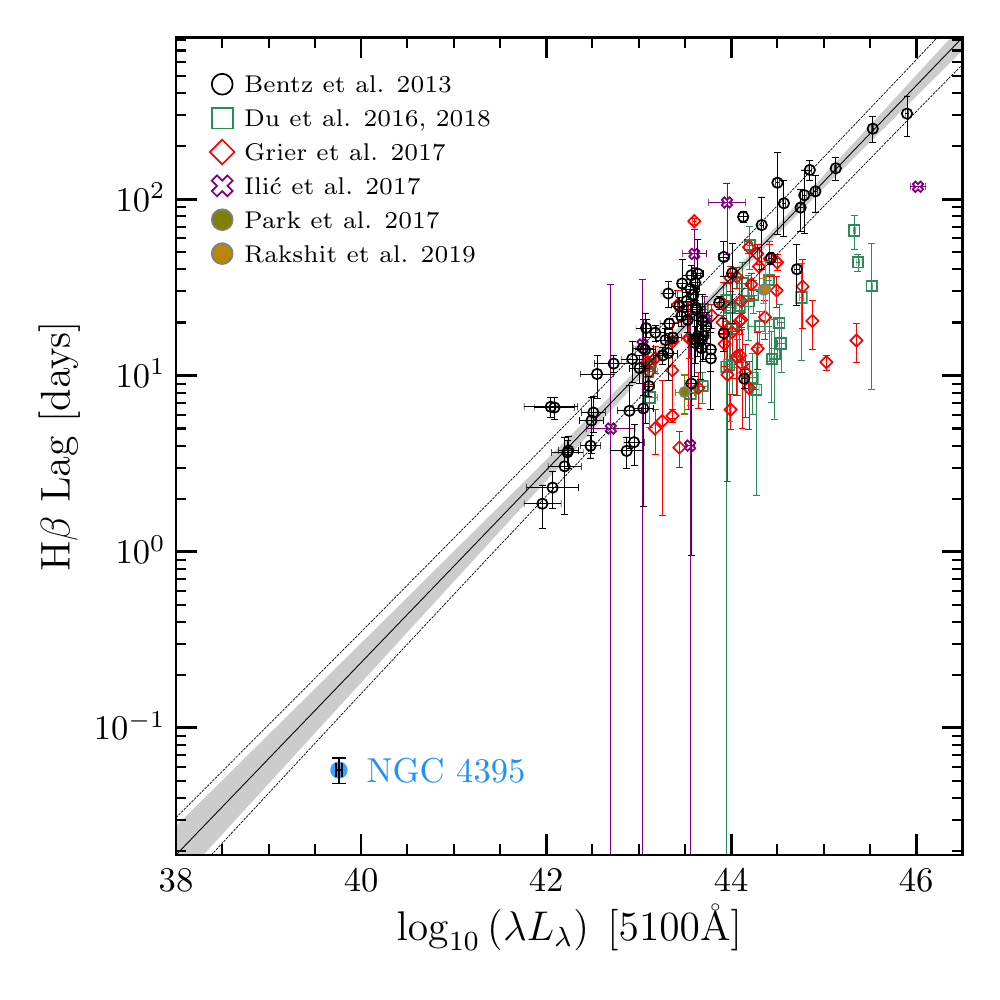}
	\caption{BLR radius vs. 5100\,\AA\ AGN luminosity. Best-fit relation by	\citet[\emph{Clean+ExtCorr} in Table 14]{Bentz+13} is shown as a black solid line, with the shaded region indicating the 1$\sigma$ confidence interval of the  fit and the dotted lines representing the 1$\sigma$ prediction interval of AGNs considering intrinsic scatter. AGNs shown here are from \cite{Bentz+13}, as well as \citet{Du+16, Du+18}, \cite{Grier+17}, \cite{Ilic+17}, \cite{Park(SY)+17}, and \cite{Rakshit+19}. NGC 4395 is shown as a blue circle with black error bar. 
		\label{fig:sl}} 
\end{figure}

\section{Discussion}

\subsection{Comparison with Previous Studies}\label{ss:ps}

There have been several previous studies of the emission-line time lag in NGC 4395, and our time-lag measurement is broadly consistent with these results. For example, \cite{Desroches+06} measured the \ha\ line lag to be \valerrud{0.06}{0.034}{0.030} days (\valerrud{86}{49}{43} min) by integrating continuum-subtracted line spectra, consistent to our measurement, while \cite{Edri+12} measured a lag of $3.6 \pm 0.8$\,hr ($216 \pm 48$ min) based on photometric light curves with broad-band filters (SDSS \emph{g'}, \emph{r'}, and \emph{i'}). In the case of the broad-band photometry light curves, they measured auto-correlations for each light curve as well as cross-correlations for each combination of two light curves, and then subtracted auto-correlations from cross-correlations to determine the lag between the continuum and emission line. However, this method is less reliable since the continuum flux is dominant ($> 75$\%) in the total flux measured with the broad-band filters, leading to the difficulty that the flux measurement is more prone to photometric errors. Lastly, \cite{Peterson+05} reported the lag between the continuum at 1350\,\AA\ and the \civ\ $\lambda 1549$ broad emission line as $\sim 1$\,hr based on two different sets of light curves. The \civ\ lag is shorter than our \Ha\ lag, indicating that these measurements are consistent with the stratification of the BLR. 

In the case of AGN luminosity, \cite{Filippenko&Ho03} reported $L_{5100} = 6.6\times 10^{39}\,\mathrm{erg\,s^{-1}}$, which is close to our estimate. Other studies determined the bolometric luminosity of NGC 4395 by integrating the SED as  $1.2\times 10^{41}\,\mathrm{erg\,s^{-1}}$ \citep{Lira+99}, $1.9\times 10^{40}\,\mathrm{erg\,s^{-1}}$  \citep{Moran+99}, and $9.9\times 10^{40}\,\mathrm{erg\,s^{-1}}$ \citep{Brum+19}. If we adopt the bolometric correction of 10 \citep{Woo&Urry02} for the reported measurements, $L_{\rm 5100}$ ranges from $1.9\times10^{39}\,\mathrm{erg\,s^{-1}}$ to $1.2\times 10^{40}\,\mathrm{erg\,s^{-1}}$, which are similar to our estimate. 

The current investigation along with prior studies reporting estimates of the AGN luminosity are affected by various sources of systematic uncertainties. Note that the AGN PSF decomposition using high-quality imaging data has been applied to many of the reverberation-mapped AGNs \citep{Bentz+13}. However, even the best spatial resolution of {\it Hubble Space Telescope (HST)} imaging may not be enough to reliably separate the AGN from the NSC with an effective radius of $< 0.3$\arcsec. 

Although the luminosity of the AGN in NGC 
4395 is two orders of magnitude lower than in typical Seyfert 1 galaxies, the AGN seems to broadly follow the size-luminosity relation, indicating that the same photoionization assumption is valid at the low-luminosity end. Based on the black hole mass measurement from the \ha\ reverberation mapping \citep[9000 \msun ,][]{Woo+19}, and the bolometric luminosity log $L_{\rm bol} = 42.06$, we determined the Eddington ratio to be $\sim 5$\%. These results indicate that NGC 4395 is a scaled-down version of a typical Seyfert 1 galaxy with an intermediate-mass black hole and $\sim 5$\% of the Eddington accretion.

\subsection{Variability}

We measured the variability of the AGN continuum in the V-band as $\fvar\approx 0.02$ and $R_\mathrm{max}\approx 1.1$ based on one-day baseline light curves. The amplitude of the variability slightly increases as $\fvar\approx 0.04$--0.08 and $R_\mathrm{max}\approx 1.2$--1.7 with a longer baseline of several days. These results are consistent with those of \cite{Desroches+06}, who reported the variability in V-band photometry $\fvar=0.019$--0.042 and $R_\mathrm{max}=1.08$--1.20 based on single-night light curves. The variability of NGC 4395 is similar to those of other Seyfert 1 galaxies \citep[e.g.,][]{Walsh+09}. For example, the 15 AGNs with relatively low luminosity, which were monitored by the Lick AGN Monitoring Project 2011 over several-month timescales, showed $\sim 0.1$\,mag variability, $\fvar$ ranges of 0.02--0.13, and $R_\mathrm{max}$ ranges of 1.13--1.68 \citep{Pancoast+19}. These results imply that the variability characteristics of NGC 4395 are similar to those of other Seyfert 1 galaxies. 

\subsection{The Size-Luminosity Relation of NGC 4395}\label{ss:sl}
We investigated the size-luminosity relation at the low-luminosity end by including our lag and luminosity measurements of NGC 4395. While the size-luminosity relation has been defined based on the \hb\ lag measurements, we only obtained an \ha\ lag measurement. Thus, the systematic difference between \Hb\ and \Ha\ lags may introduce additional uncertainty. 

It is not clear whether the \ha\ time lag is longer than the more commonly used \hb\ lag for a given object. \cite{Kaspi+00} found no significant difference between continuum-to-\ha\ and continuum-to-\hb\ time-lag measurements in their reverberation sample. In contrast, \ha\ is expected to show a longer time lag than \hb\ owing to optical-depth effects, which are manifested as the radial stratification within the BLR \citep{Netzer75, Rees+89, Korista&Goad04}. \cite{Bentz+10} provided a detailed discussion, reporting that the \Ha\ lag is a factor of 1.54 longer on average than the \hb\ lag based on the reverberation-mapping results of low-redshift AGNs. If we assume that the \hb\ lag is shorter than the \ha\ lag, the offset of NGC 4395 from the size-luminosity relation becomes larger.

On the other hand, we need to consider the uncertainty of the measured AGN luminosity. The main systematic uncertainty comes from the correction for the flux from the NCS, which is not easily decomposed from the AGN. We adopted the luminosity of the NSC measured by \cite{Carson+15}, which suffers large uncertainty due to the limited spatial resolution. Note that the AGN and the NSC have comparable effective radii, and even with the spatial resolution provided by {\it HST}, the two sources were not clearly decomposed in the two-dimensional imaging analysis. \cite{Carson+15} argued that the degeneracy between the NSC and the AGN introduced a systematic uncertainty of 0.2\,mag for the luminosity of the NSC. We note that based on our high-quality GMOS spectrum we were not able to decompose the AGN power-law component and stellar component. Considering the degeneracy of the AGN and the NSC in the imaging and spectroscopy and the dependence of the flux ratio on wavelength, the overall uncertainty of the AGN luminosity seems considerable. Thus, we find no strong evidence that NGC 4395 is offset from the size-luminosity relation defined by more-luminous AGNs. 

Given the measured luminosity and size of the BLR, the offset of NGC 4395 from the size-luminosity relation is not significantly large when compared to more recent reverberation-mapping results. For example, AGNs from the studies by Du et al. (\citeyear{Du+16}, \citeyear{Du+18}), \cite{Grier+17}, and \cite{Ilic+17} exhibit large scatter, and some of them are more offset than NGC 4395. Note that the AGNs in Du et al. (\citeyear{Du+16}, \citeyear{Du+18}) have a high accretion rate (e.g., super-Eddington), which may be the reason for the offset from the relation. In contrast, NGC 4395 has a much lower Eddington ratio ($\sim 5$\%). On the other hand, the AGNs studied by \cite{Grier+17} are higher-redshift objects with Eddington ratio larger than 0.1, and their luminosity may suffer systematic uncertainties due to the contribution from the host galaxies. To constrain the size-luminosity relation at the low-luminosity end, it is necessary to obtain better measurements of the AGN luminosity of NGC 4395 and to investigate the scatter of the relation caused by systematic effects and Eddington ratios.

\section{Summary and Conclusions}

We present observations of the variability of NGC 4395 along with reverberation-mapping results using the photometric data from our monitoring campaign, which consisted of optical observations during 5 nights in 2017 and 3 nights in 2018, UV observations during 4 nights in 2017, and near-IR observations during 3 nights in 2017. 

In 2017, we measured the variability in the V band as $\sigma_{m,\,\mathrm{V}}=0.10$, $R_\mathrm{max,\, V}=1.70$, and $\fvar_\mathrm{,\,V}=0.082$, while in 2018, we measured $\sigma_{m,\,\mathrm{V}}=0.05$, $R_\mathrm{max,\, V}=1.23$, and $\fvar_\mathrm{,\,V}=0.047$. In 2017, we measured the fractional RMS variability for the UVM2, J, H, and K bands and observed a decreasing trend with increasing wavelength, shown as  $\fvar_\mathrm{,\,UVM2}=0.081$, $\fvar_\mathrm{,\,V}=0.082$, $\fvar_\mathrm{,\,J}=0.027$, $\fvar_\mathrm{,\,H}=0.016$, and $\fvar_\mathrm{,\,K}=0.011$. Based on the single-night light curves, we measured the variability of the V band and \ha\ to be $\sigma_{m,\,\mathrm{V}}=0.03$, $R_\mathrm{max,\, V}=1.11$, $\fvar_{,\, \mathrm{V}}=0.025$, $\sigma_{m,\,\mathrm{H\alpha}}=0.01$, $R_\mathrm{max,\, H\alpha}=1.04$, and $\fvar_{,\, \mathrm{H\alpha}}=0.001$ on 2017-05-02, and $\sigma_{m,\,\mathrm{V}}=0.02$, $R_\mathrm{max,\, V}=1.10$, $\fvar_{,\, \mathrm{V}}=0.017$, $\sigma_{m,\,\mathrm{H\alpha}}=0.01$, $R_\mathrm{max,\, H\alpha}=1.03$, and $\fvar_{,\, \mathrm{H\alpha}}=0.006$ on 2018-04-08. 

We performed the cross-correlation analysis using various pairs of light curves. For the time lag of the \Ha\ emission line with respect to the V-band continuum, we demonstrated that the correction for the continuum in the narrow \Ha-band filter significantly changed the lag since the variability of the continuum flux correlates with that of the V band. Without a proper correction for the continuum contribution, the lag can be underestimated by a factor of $\sim 2$. Our best estimate for the \ha\ lag measurement is \valerrud{83}{13}{14} min, which is based on the light curves from 2018-04-08 after correcting for the continuum contribution in the narrow \Ha-band by assuming the variability amplitude is the same as that of the V-band light curve. The \Ha\ lag measurements from other light curves from various nights are consistent with the best lag measurement, although these measurements are much less reliable owing to much lower quality light curves. In the case of the UV-to-V, and among near-IR bands (J, H, K), we did not find a reliable lag measurement. 

We determined the monochromatic luminosity at 5100\,\AA\ of the AGN in NGC 4395 by analyzing the best imaging data and the mean spectrum from the Gemini GMOS. However, the central point source also includes the flux from a nuclear star cluster. Thus, this luminosity is an upper limit. By subtracting an estimate of the luminosity of the nuclear star cluster, we obtained $\log_{10}\lumcont/[\mathrm{erg\,s^{-1}}]=39.76\pm0.03$, which is two orders of magnitudes lower than that of any Seyfert 1 galaxy with available reverberation mapping results. 

We investigated the size-luminosity relation of NGC 4395 by comparing with more luminous Type 1 AGNs, finding that the relation extends to very low luminosity. This result suggests that the naive photoionization expectation is valid in this low-luminosity regime. While NGC 4395 has a very low AGN luminosity, the Eddington ratio of NGC 4395 is $\sim0.05$, indicating that this AGN is similar to typical Seyfert 1 galaxies. Nevertheless, the offset of NGC 4395 from the best-fit size-luminosity relation of \citet{Bentz+13} is significant by 0.48\,dex ($\geq 2.5\sigma$), indicating that the extrapolation of the previously defined size-luminosity relation down to intermediate-mass black holes, or low-luminosity AGNs, would introduce a large systematic uncertainty in black hole mass estimates. The systematic uncertainty of the size-luminosity relation has been already noted by more recent reverberation studies \citep[e.g.,][]{Du+16, Du+18, Grier+17}. In order to define the low-luminosity end of the size-luminosity relation, it is necessary to perform reverberation analysis for additional targets with similar luminosities.  

NGC 4395 is a unique object with an intermediate black hole mass ($\sim10^{4}$ \msun) and may have similar properties compared to typical Seyfert 1 galaxies. Further investigation of the detailed AGN properties, such as the X-ray SED and gas outflows, will shed light on the understanding of intermediate-mass AGNs.  

\acknowledgements{
This work has been supported by the Basic Science Research Program through the National Research Foundation of Korean Government (2016R1A2B3011457), and by Samsung Science and Technology Foundation under Project Number SSTF-BA1501-05. The work of H.C. was supported by an NRF grant funded by the Korean Government (NRF-2018H1A2A1061365-Global Ph.D. Fellowship Program).
A.M.T., N.I.Sh., A.E.N., and V.L.O. thank the Program of Development of the Lomonosov Moscow State University. 
V.L.O. thanks Prof. Ari Laor for useful discussions and interest in this project. D.I., A.K., L.C.P., and O.V. gratefully acknowledge the support of the Ministry of
Education, Science and Technological Development of the Republic of Serbia
through projects 176001, 176004, 176005, and 176021.
A.V.F.'s group was financially supported by
the TABASGO Foundation, the Christopher R. Redlich Fund, and the
Miller Institute for Basic Research in Science (U.C. Berkeley).
}

\bibliography{bib}{}

\end{document}